\documentclass[fleqn,usenatbib]{mnras}

\usepackage[T1]{fontenc}
\usepackage{ae,aecompl}

\usepackage{newtxtext}
\usepackage{newtxmath}

\usepackage{ragged2e}

\usepackage{graphicx}
\usepackage{amsmath}
\usepackage{amssymb}
\usepackage{bm}

\usepackage[usenames,dvipsnames]{xcolor}
\usepackage{tikz}

\providecommand{\sorthelp}[1]{}

\title[Joint ALMA-Bolocam-{\it Planck} study of RXJ1347]{A joint ALMA-Bolocam-{\it Planck} SZ study of the pressure distribution in RX~J1347.5--1145}
\author[L. Di Mascolo et al.]{
  Luca Di Mascolo$^{1}$\thanks{E-mail: lucadim@mpa-garching.mpg.de}, 
  Eugene Churazov$^{1,2}$, 
  Tony Mroczkowski$^{3}$\\
  $^{1}$Max-Planck-Institut f\"{u}r Astrophysik, Karl-Schwarzschild-Strasse 1, Garching D-85741, Germany\\
  $^{2}$Space Research Institute, Profsoyuznaya 84/32, Moscow 117997, Russia\\
  $^{3}$European Southern Observatory, Karl-Schwarzschild-Strasse 2, Garching D-85748, Germany
}

\date{Accepted 2019 May 31. Received 2019 May 29; in original form 2018 December 03}
\pubyear{2019}

\begin{document}
\label{firstpage}
\pagerange{\pageref{firstpage}--\pageref{lastpage}}
\maketitle

\begin{abstract}
We report the joint analysis of single-dish and interferometric observations of the Sunyaev--Zeldovich (SZ) effect from the galaxy cluster RX~J1347.5--1145. We have developed a parametric fitting procedure that uses native imaging and visibility data, and tested it using the rich data sets from ALMA, Bolocam, and {\it Planck} available for this object. RX~J1347.5--1145 is a very hot and luminous cluster showing signatures of a merger. Previous X-ray-motivated SZ studies have highlighted the presence of an excess SZ signal south-east of the X-ray peak, which was generally interpreted as a strong, shock-induced pressure perturbation. Our model, when centred at the X-ray peak, confirms this. However, the presence of two almost equally bright giant elliptical galaxies separated by $\sim100\;{\rm kpc}$ makes the choice of the cluster centre ambiguous, and allows for considerable freedom in modelling the structure of the galaxy cluster. For instance, we have shown that the SZ signal can be well-described by a single smooth ellipsoidal generalized Navarro--Frenk--White profile, where the best-fitting centroid is located between the two brightest cluster galaxies. This leads to a considerably weaker excess SZ signal from the south-eastern substructure.  Further, the most prominent features seen in the X-ray can be explained as predominantly isobaric structures, alleviating the need for highly supersonic velocities, although overpressurized regions associated with the moving subhaloes are still present in our model.
\end{abstract}


\section{Introduction}\label{sec:introduction}
The Sunyaev--Zeldovich (SZ) effect observed in the direction of a galaxy cluster is a spectral distortion of the cosmic microwave background \citep[CMB;][]{Sunyaev1972}. The dominant contribution is given by the renowned thermal SZ effect, and is induced by the inverse Compton scattering of CMB photons off the thermal population of free electrons within the intracluster medium \citep[ICM; for a recent review, see][]{Mroczkowski2019}. The thermal SZ effect directly measures the line-of-sight integral of the electron thermal pressure, and is complementary to X-ray observations, which map the bremsstrahlung emission of thermal electrons colliding with ions. The X-ray surface brightness is therefore proportional to the integral of the product of electron and ion densities, while the electron temperature can be determined through X-ray spectroscopy \citep[for a review,][]{Sarazin1986}. Thus, in combination with X-ray measurements, the SZ effect can provide a powerful tool for probing the internal structure of the ICM, which is continuously perturbed by the accretion and merger events arising as a result of the hierarchical clustering, as well as feedback processes from active galactic nuclei \citep[AGNs; see e.g.][]{Kravtsov2012}. It follows then that the analysis of the resulting deviations from the global thermodynamic properties can yield insights into the physics and thermodynamics of galaxy clusters and their formation history.

In recent years, the SZ effect has been observed at increasingly higher angular resolution and sensitivity, allowing observers to obtain information on spatial scales comparable to those probed by X-ray measurements. Indeed, instead of being limited solely to unresolved or poorly resolved detections, it has opened a millimetre-wave window on the nature and properties of the small-scale substructures in the ICM. High-resolution measurements of the SZ effect by single-dish facilities such as MUSTANG-2 on the 100-meter Green Bank Telescope, with a full width at half maximum (FWHM) beam size of $9~\mathrm{arcsec}$ at $90~\mathrm{GHz}$ \citep{Dicker2014}, or the NIKA2 instrument on the IRAM 30-meter telescope, with $17.5~\mathrm{arcsec}$ and $11~\mathrm{arcsec}~\mathrm{FWHM}$ beams respectively at $150~\mathrm{GHz}$ and $260~\mathrm{GHz}$ \citep{Adam2018a}, now allow the characterization of pressure substructure in a growing number of clusters. In particular, both parametric (e.g. \citealt{Adam2014}, \citealt{Romero2015}) and non-parametric methods (e.g. \citealt{Ruppin2017}, \citealt{Romero2018}) have proven to provide a reliable description of the ICM pressure profile, while structure-enhancement filtering techniques have been successfully applied for detecting discontinuities in cluster SZ surface brightness maps \citep{Adam2018b}.

Nevertheless, radio interferometers remain the only instruments so far capable of measuring the SZ effect with an angular resolution better than $5~\mathrm{arcsec}$\!. The first pioneering observations of the SZ effect with the Atacama Large Millimeter/Submillimeter Array (ALMA) demonstrated its ability to provide images of galaxy clusters with such unprecedented angular resolution \citep{Kitayama2016} and to detect discontinuities in the pressure distribution of the ICM \citep{Basu2016}. Unfortunately,  analysis techniques used for single-dish SZ effect imaging cannot be easily generalized for the fitting of images obtained from interferometric data. Among the main issues is the introduction of non-trivial correlation in the image noise. This is induced by the intrinsic non-linearity of the deconvolution methods \citep[e.g. the standard CLEAN approach by][]{Hoegbom1974} employed for recovering information about the Fourier modes of the sky signal that are not sampled in a given interferometric observation. Moreover, any chosen weighting and gridding schemes, along with the specific image-reconstruction technique, may possibly introduce artefacts in the deconvolved image, and therefore bias its interpretation. 
Additionally, incomplete sampling of the Fourier domain causes the so-called missing-flux issue, which arises from the impossibility of constraining scales larger than those corresponding to spacings smaller than the minimum baseline length. Several techniques --- feathering, deconvolution informed by total flux measurements, synthetic short-spacing (we refer the reader to \citealt{Stanimirovic2002} for an overview) --- have been developed to include short-spacing information at different steps of the interferometric imaging process, but these still rely on highly non-linear deconvolution algorithms. It is therefore clear that a straightforward solution may be to fit the visibility data directly in Fourier space through a forward-modelling procedure.  The visibility data exhibit nearly Gaussian noise, and modelling in Fourier space allows for full knowledge of the instrument sampling function. In order to fully exploit the potentialities of visibility modelling, a range of tools have been developed (see e.g. \texttt{uvmultifit} by \citealt{MartiVidal2014} or \texttt{galario} by \citealt{Tazzari2018}). Indeed, analogous approaches have already been shown over the past two decadeds to provide a reliable technique for studying interferometric observations of the SZ effect \citep[see e.g.][for an incomplete list of examples of applications of the interferometric modelling technique]{Carlstrom1996,LaRoque2006,Feroz2009,Mroczkowski2009,Basu2016,Abdulla2019}. We therefore follow this method as well in our joint SZ analysis, rather than relying on image-space techniques, while incorporating some of the recent advances in Bayesian and $uv$-space modelling.

In fact, the short-spacing problem is particularly relevant for sources covering large fractions of, or extending beyond, the field of view of the instrument. In the case of galaxy clusters, this manifests itself in a significant high-pass spatial filtering of the extended SZ effect signal. Additional large-scale constraints are required in order to correctly derive a global description of the pressure distribution in galaxy clusters. Several studies \citep[see e.g.][]{Romero2018} have already shown the importance of joint analysis of both low- and high-resolution observations when attempting to obtain information over a broad range of spatial scales. In general, all previous studies of the SZ effect we are aware of that have combined SZ data from instruments sampling different spatial frequencies have either been limited to image-space or to interferometric SZ data exclusively.\footnote{We note however that there have been several studies over the past two decades relying on joint likelihood analyses of X-ray surface brightness imaging data with interferometric SZ observations \citep[e.g.][]{Reese2000,LaRoque2006,Mroczkowski2009}.} In this work, we extend the analysis for combining both interferometric and single-dish measurements by modelling the thermal SZ effect signal from galaxy clusters through the joint fitting of SZ imaging and interferometric data.

\begin{figure}
    \begin{center}
    \includegraphics[width=\columnwidth]{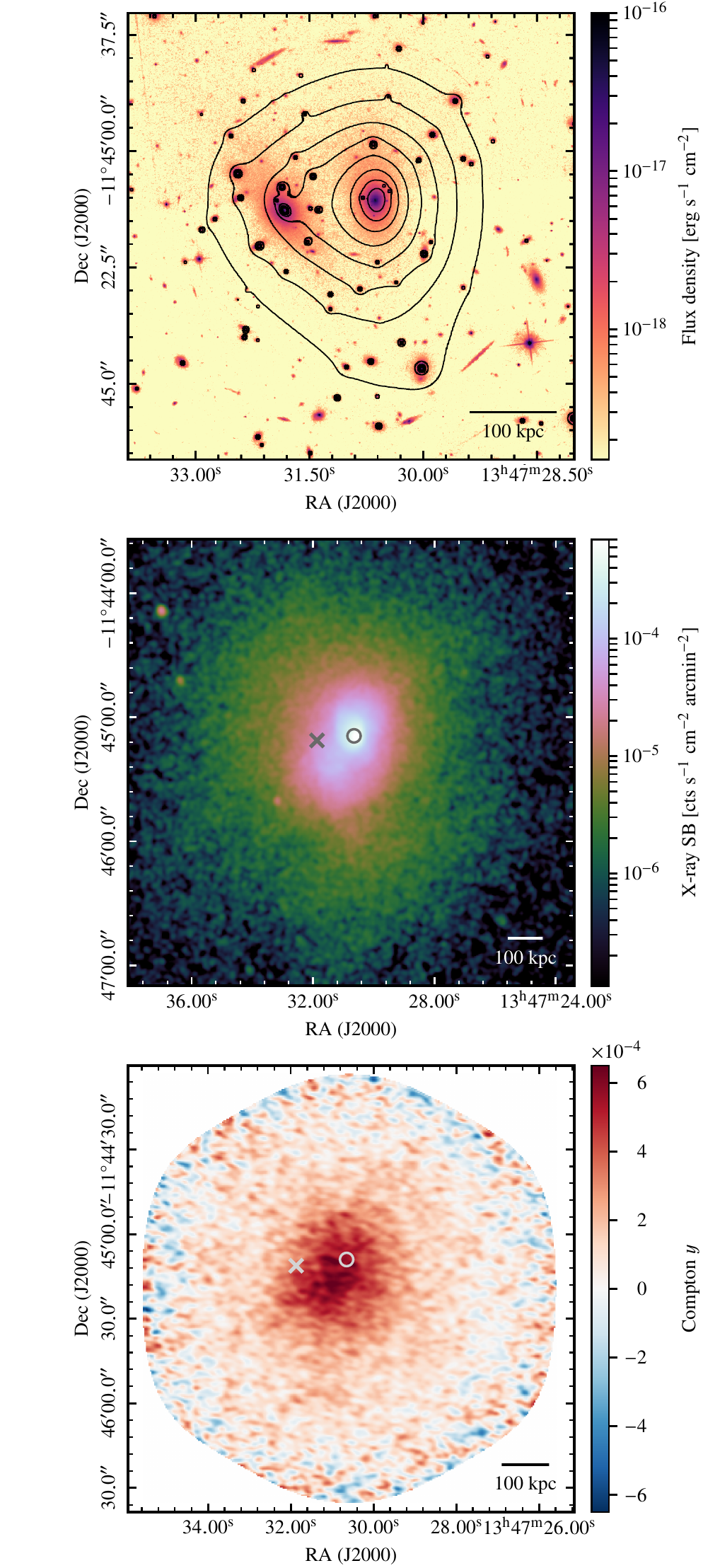}
    \end{center}
    \caption{Multi-wavelength view of the galaxy cluster RX J1347.5-1145. Shown are the HST-ACS optical image (top), the {\it Chandra} 0.5-3.5 keV X-ray surface brightness (SB) map (middle), and the Compton $\vary$ image of the SZ effect created by feathering the ACA, ALMA, and Bolocam data (bottom; see Section~\ref{sec:data} for a description of the SZ data). Overlaid in the top panel are the contours from the \citet{Zitrin2015} light-traces-matter lensing $\kappa$-map. The circle and the cross in the X-ray and SZ images indicate respectively the positions of the western (wBCG; $\mathrm{13^h 47^m 30\fs650,-11^d 45^m 09\fs00}$) and eastern (eBCG; $\mathrm{13^h 47^m 31\fs870,-11^d 45^m 11\fs20}$) of the two dominant cluster galaxies observed in the optical map. Note that each panel has a different scale and centre.}
    \label{fig:figure01}
\end{figure}

As a test case, we apply our joint image-visibility model reconstruction technique to single-dish and interferometric observations of the SZ effect from the galaxy cluster RX~J1347.5--1145 ($z=0.451$). It is among the most massive and X-ray luminous clusters ever observed, which have made it the ideal target of observations over a broad range of wavelengths (Fig.~\ref{fig:figure01}). In particular, due to the availability of a number of millimetre measurements of the SZ effect in the direction of the cluster, covering several frequencies and spatial resolutions, it provides an excellent test bed for probing the applicability of the combined study of visibilities and single-dish data. 

The paper is organized as follows. In Section~\ref{sec:rxj1347}, we give a brief summary of previous observations and studies of RX~J1347.5--1145. We then provide an overview of the data employed in our analysis in Section~\ref{sec:data}. A description of the analysis of the SZ signal from RX~J1347.5--1145 and the reconstruction of its pressure profile are presented in Sections~\ref{sec:analysis} and ~\ref{sec:results}, respectively. The latter also discusses the physical and thermodynamic interpretation of RX J1347.5--1145, comparing the results from our SZ modelling with an independent X-ray analysis. In Section~\ref{sec:conclusions}, we summarize and offer concluding remarks.

Throughout this work, we assume a flat $\Lambda$CDM cosmology with $\Omega_m=0.308$, $\Omega_{\Lambda}=0.692$, and $H_0=67.8~\mathrm{km}~\mathrm{s}^{-1}~\mathrm{Mpc}^{-1}$, consistent with the {\it Planck} cosmological results \citep{Planck2015XIII}. Within this cosmology, 1~arcsec corresponds to a physical scale of 5.94 $\mathrm{kpc}$ at the redshift of RX~J1347.5--1145.

\section{RX~J1347.5--1145}\label{sec:rxj1347}
RX~J1347.5--1145 is a massive galaxy cluster discovered in the ROSAT X-ray all-sky survey \citep{Schindler1995,Voges1999}. The studies based on the low angular resolution ROSAT X-ray data initially highlighted a spherical, strongly peaked surface brightness profile, suggesting the cluster to be dynamically relaxed and characterized by a cool central region \citep{Schindler1997}. However, the high-resolution measurements of the SZ effect in the direction of RX~J1347.5--1145 performed by the Nobeyama Bolometer Array \citep[NOBA;][]{Komatsu2001} provided early indications of a significant enhancement of the SZ signal to the south-east of the X-ray emission peak (i.e.\ the 'south-eastern SZ excess'). Subsequent X-ray observations of the cluster by {\it Chandra} \citep{Allen2002}, {\it XMM-Newton} \citep{Gitti2004}, and {\it Suzaku} \citep{Ota2008} confirmed the existence of a south-eastern extension in the proximity of the core region, manifesting temperatures higher than the average value of the surrounding ICM. The evidence of a disturbed SZ morphology was further supported by the high-resolution SZ imaging of both single-dish observations \citep{Mason2010,Adam2014} and radio interferometric data \citep{Plagge2013,Kitayama2016}, which additionally allowed for identifying a potential pressure discontinuity east of the X-ray peak \citep{Mason2010,Adam2018b}. The analyses of, for example, \cite{Korngut2011} and \cite{Plagge2013} further determined that the excess could account for $\sim9-10\%$ of the total thermal energy of the cluster, assuming the bulk pressure distribution of the cluster can be described by a spherically-symmetric model. 

The current interpretation of the observed cluster morphology relates the south-eastern structure to gas that has been stripped away and shock-heated as a consequence of a major merging event. In this scenario, the involved subcluster is assumed to be moving in the south-west-north-east direction and to strongly perturb the main, initially-relaxed, cool-core cluster component. This is also consistent with the results of the weak- and strong-lensing analyses of optical data \citep{Bradac2008,Koehlinger2014,Zitrin2015,Ueda2018}, which show that the projected mass density has a primary peak centred near the AGN embedded in the cool core, and an additional component elongated towards a secondary peak at or near the brightest cluster galaxy (BCG) to the east of the X-ray peak (hereafter `eBCG' to distinguish it from that coincident with the cluster core, which we refer to as `wBCG'; see the top panel of Fig.~\ref{fig:figure01}, which includes the lensing contours from \citealt{Zitrin2015}). Furthermore, optical spectroscopic analysis constrains the dynamics of the merger to take place mainly in the plane of the sky \citep{Miranda2008,Lu2010}. This is corroborated by the small difference in the redshifts of the two dominant BCGs, measured to be of the order of $\sim 100~\mathrm{km s^{-1}}$ \citep{Cohen2002}. On the other hand, a radio mini-halo has been detected in the direction of the cool-core region \citep{Gitti2007,Ferrari2011}, and has been considered as an indication of the possible occurrence of sloshing gas within the cluster core. In fact, diffuse radio emission has been found to be spatially correlated with the cold fronts generated by the sloshing gas motions \citep{Mazzotta2008,Zuhone2013}.
However, although the comparison of the observed X-ray surface brightness and hydrodynamic simulations further favours the scenario of the sloshing gas and south-eastern substructure as due to a plane-of-sky merger, the cluster merger dynamics and geometry are still subjects of debate \citep{Johnson2012,Kreisch2016}. More recently, \citet{Ueda2018} combined X-ray, strong-lensing and interferometric SZ observations to study RX~J1347.5--1145. Along with confirming the correspondence of the SZ enhancement with stripped gas that has been shock-heated to high temperatures, they also reported that the sloshing in the cluster core seen in X-ray data is not accompanied by large pressure variations, suggesting subsonic gas velocities in this region. The compact structure of the characteristic spiral pattern observed in the cool-core region has been considered as an indication that it has been plausibly induced by a secondary, minor interaction instead of the major merger related to the south-eastern substructure.

\section{Data overview}\label{sec:data}
Here we present the set of single-dish and interferometric observations of employed in our joint analysis. A summary the observations can be found in Table~\ref{tab:data1}.

\begin{table*}
    \caption{Details of the observations used for modelling RX~J1347.5--1145.}
    \label{tab:data1}
    \begin{center}
    \begin{tabular}{lcccccl}
    \hline
    Telescope & Average RMS & Resolution & Largest scale & FoV & Frequency & Reference\\
    \hline
    ALMA$^a$   & $12~\mathrm{\mu Jy~beam^{-1}}$ & $( 4.1, 2.4)~\mathrm{arcsec}$ & $58.8~\mathrm{arcsec}$ &  $62~\mathrm{arcsec}$ & $84$-$88~\mathrm{GHz}$, $96$-$100~\mathrm{GHz}$  & \cite{Kitayama2016}\\
    ACA$^a$    & $83~\mathrm{\mu Jy~beam^{-1}}$ & $(20.5,11.1)~\mathrm{arcsec}$ & $99.7~\mathrm{arcsec}$ & $107~\mathrm{arcsec}$ & $84$-$88~\mathrm{GHz}$, $96$-$100~\mathrm{GHz}$  &\cite{Kitayama2016} \\
    Bolocam & $0.38~\mathrm{mJy~beam^{-1}}$ & $ 58~\mathrm{arcsec}$ & $8.9~\mathrm{arcmin^b}$ & $14~\mathrm{arcmin}$ & $140~\mathrm{GHz}$ & \cite{Sayers2013}\\
    {\it Planck}  & $1.2\cdot10^{-6}$ [Compton $\vary$] & $ 10~\mathrm{arcmin}$ & Full-sky & Full-sky & ---$^c$ & \cite{Planck2015I}\\
    \hline
    \end{tabular}
    \end{center}
    \vspace{-5pt}
    \justifying{
    $^a$ The noise RMS reported for the interferometric measurements is the average noise level measured from the dirty images generated by adopting a natural weighting scheme, while the resolution is provided in terms of the FWHM major and minor axes of the resulting synthesized beam. The largest scale is measured as the inverse of the shortest array baseline in units of wavelengths. \\
    $^b$ The largest mode recovered by Bolocam is the spatial scale corresponding to the transfer function HWHM frequency.\\
    $^c$ Instead of considering single-frequency images, we employ the MILCA Compton $\vary$ map generated by combining all the {\it Planck} HFI data (100-857 GHz).\\}
\end{table*}

\subsection{Atacama Large Millimeter Array}\label{ssec:alma}
\begin{figure}
  \begin{center}
  \includegraphics[width=\columnwidth]{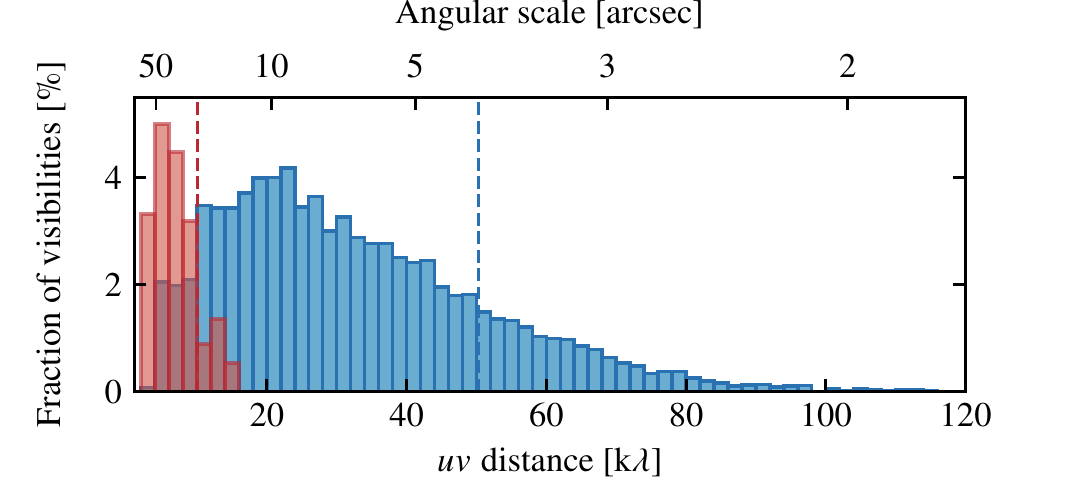}\\\vspace{0.5cm}
  \includegraphics[width=\columnwidth]{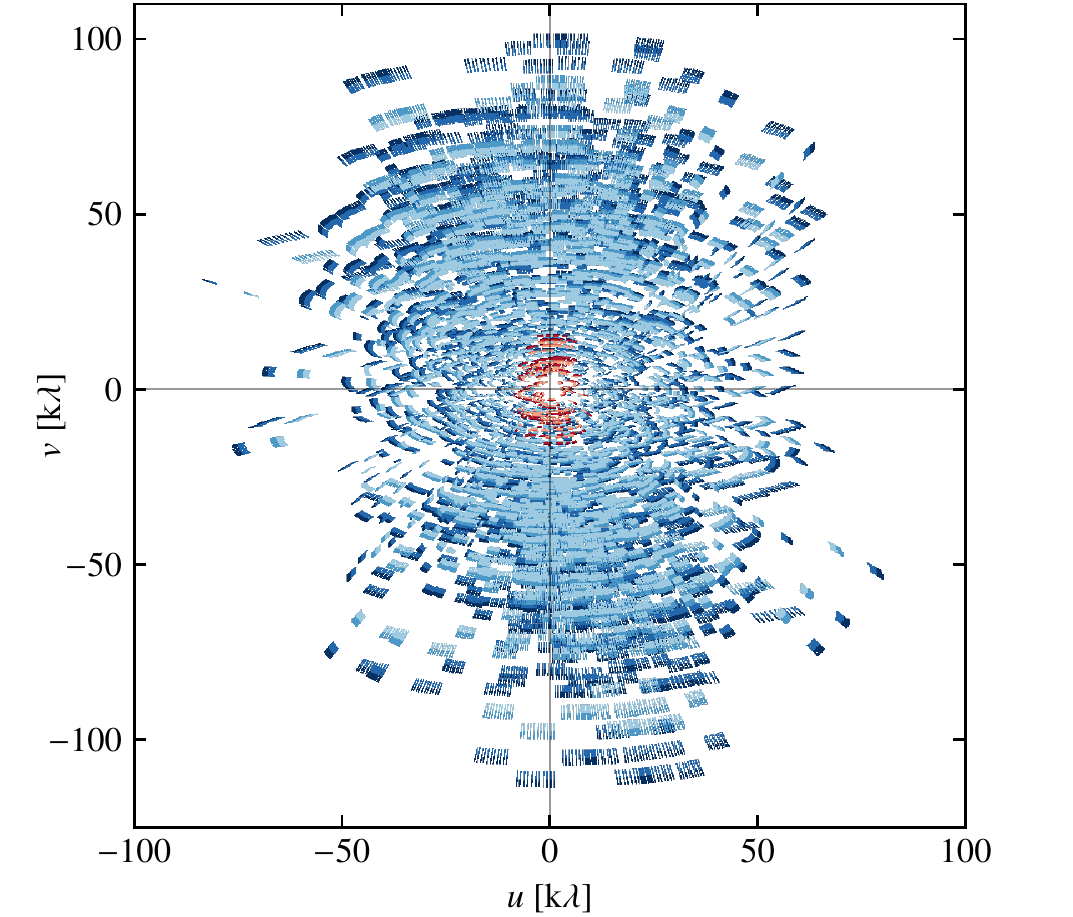}
  \end{center}
  \caption{Histogram of the sampled $uv$ distances (top) and coverage of the $uv$-plane (bottom) for the interferometric observation of RX~J1347.5--1145 with the ALMA 12-meter (blue) and ACA 7-meter (red) arrays. The different shades of each colour indicate separate spectral windows. The vertical lines in the top panel refer to the major axis FWHM of the synthesized beams obtained by imaging the ACA and ALMA data separately, and adopting natural weights. We computed the fraction of visibilities per bin in $uv$ distance with respect to the cumulative number of both ACA and ALMA data points.}
  \label{fig:figure03}
\end{figure}

RX~J1347.5--1145 was observed by both the main ALMA $12\mathrm{m}$ array and $7\mathrm{m}$ Atacama Compact Array (ACA; a.k.a. Morita Array) during Cycle 2. The galaxy cluster was mapped employing seven mosaic pointings, each using four 2~GHz wide spectral windows.  The spectral windows were centred at $85$, $87$, $97$, and $99~\mathrm{GHz}$, that is in ALMA Band 3, which ranges 84-116~GHz. The combination of the two arrays resulted in visibilities covering the $uv$-plane between $2.1$ and $115.9~\mathrm{k}\lambda$, corresponding to spatial scales of $1.66~\mathrm{arcmin}$ and $1.78~\mathrm{arcsec}$, respectively (Fig.~\ref{fig:figure03}). We refer to \cite{Kitayama2016} for a more detailed description of the combined ACA+ALMA observation of RX~J1347.5--1145.

We performed manual calibration of the ACA and ALMA measurement sets using version 4.7.2 of the Common Astronomy Software Applications \citep[\textsc{casa},][]{McMullin2007}, obtaining calibrated visibilities consistent with the data presented in \cite{Kitayama2016}. We also assume an uncertainty on the overall flux calibration of $6\%$ for both the ALMA and ACA data given the variance of the measured calibrator fluxes. All the interferometric images presented in this work, which are not used for analysis, were generated with \textsc{casa} version 5.3.0.

Rather than modelling the full data sets, we consider spectral window-averaged visibilities for each of the fields. These are computed considering the weighted average of both the $uv$ coordinates and complex values of the visibility points over a set of optimal bins defined as in \citet{Hobson1995}. We assume a top-hat frequency response over each spectral window.
The primary beam model images obtained by running the CLEAN task independently for each pointing are used for accounting for primary beam attenuation when fitting the interferometric data.

The CMB multipoles corresponding to the scales probed by the ACA+ALMA observation are larger than $\ell=6750$. Above such value, the amplitude of the anisotropies intrinsic to the primary CMB is smaller than $1~\mu\mathrm{K}_{\textsc{cmb}}$. This results to be of the order of only a few percent of both the ACA and ALMA instrumental noise, even when the presence of correlated visibilities that would enhance the significance of the CMB signal is properly taken into account. Hence, we assume the CMB term in the ACA+ALMA noise covariance matrix to be negligible. On the other hand, confusion noise may still be important. Confusion from radio sources is expected to be of the order of $1~\mu\mathrm{Jy~beam^{-1}}$ for ACA and $10~\mathrm{nJy~beam^{-1}}$ for ALMA \citep[see Eq.~3.163 in][]{Condon2016}. These are below the noise levels reported in Table~\ref{tab:data1}. However, confusion due to the emission from background dusty galaxies may not be negligible. Scaling the \citet{Lindner2011} measurement of confusion at $1.1~\mathrm{mm}$ and in a $15.6~\mathrm{arcsec}$ beam (comparable to the ACA beam size), we estimate the contribution from dusty star-forming galaxies in the cosmic infrared background is $\sim 15~\mu\mathrm{Jy~beam^{-1}}$ in the ACA data. Here, we are assuming a dust emissivity spectral index of -2.5. For the ALMA 12-meter data, if we conservatively assume the sources are uncorrelated \citep[see][for discussion]{Bethermin2017}, we estimate the CIB contribution to be $\sim 1~\mu\mathrm{Jy~beam^{-1}}$. Therefore, any correlation between the data introducing off-diagonal components in the individual blocks of the noise covariance matrix is subdominant with respect to the instrumental noise. For simplicity, we then assume the ACA+ALMA block of the noise covariance matrix to be diagonal, and assign a weight to each point of the visibility function corresponding to the spectral window average of the theoretical post-calibration weights \citep{Wrobel1999}.

\subsection{Bolocam}\label{ssec:bolocam}
We complement the ACA+ALMA data with the publicly available Bolocam observation of RX~J1347.5--1145\footnote{\url{https://irsa.ipac.caltech.edu/data/Planck/release_2/ancillary-data/bolocam/}}. The 144-element bolometer array provided measurements with a resolution of $58~\mathrm{arcsec}$ at a reference frequency of $140~\mathrm{GHz}$, an uncertainty of $5\%$ on the flux calibration, and pointing accuracy to $5~\mathrm{arcsec}$. An overview of the reduction process and data products is provided in \citet{Sayers2013}.
Along with the map of RX~J1347.5--1145, the data products comprise a set of 1000 realizations of the $140~\mathrm{GHz}$ astronomical sky, including contributions from both the CMB and unresolved, point-like sources. We used them for computing the generalized covariance matrix of the Bolocam noise to be adopted in the computation of the likelihood function.

\subsection{Planck}\label{ssec:planck}
Supplementary information about the large-scale morphology of RX~J1347.5--1145 can be inferred from the {\it Planck} data.

Instead of modelling each frequency map separately, we extracted $2^\circ \times 2^\circ$ cutouts patches from all the {\it Planck} High Frequency Instrument (HFI) full-sky maps from the 2015 public release \citep{Planck2015I}, smoothed to an effective resolution of $10~\mathrm{arcmin}~\mathrm{FWHM}$, and used them to generate a Compton $\vary$ image of RX~J1347.5--1145. 
We applied a component separation method analogous to the modified Internal Linear Combination algorithm (MILCA) discussed in \cite{Hurier2013}. Since the map is generated under the requirements of removing the CMB contributions and minimising the variance in the reconstructed thermal SZ signal, we can consider the residual noise to be dominated by the uncertainties in the reconstructed Compton parameter $\vary$ map. Moreover, the associated noise covariance matrix is assumed to be diagonal, with elements equal to the pixel-by-pixel MILCA estimates of the residual RMS noise level. 

The {\it Planck} Compton $\vary$ map is also used for computing the cylindrically integrated Compton parameter $Y_{\mathrm{cyl}}$ over a solid angle up to an angular radius of $15~\mathrm{arcmin}$. We obtain 
\begin{equation}
    Y_{\mathrm{cyl}}(15~\mathrm{arcmin}) = (3.24\pm0.54)\times10^{-3}~\mathrm{arcmin}^2 \,,
    \label{eq:planck1}
\end{equation}
where the uncertainties are obtained as the RMS of the same integral computed at random positions around the galaxy cluster \citep{Adam2015}. We compare this value to the one computed by integrating over the model Compton $\vary$ map properly smoothed to the $10~\mathrm{arcmin}~\mathrm{FWHM}$ resolution of {\it Planck}.

\section{Analysis technique}\label{sec:analysis}
Our modelling tool makes use of a standard Bayesian approach to the parametric forward-modelling of the thermal SZ signal jointly from images and visibilities. It is specifically designed to allow for the simultaneous reconstruction of an arbitrary number of independent model components and corresponding spectral features by means of a multi-frequency analysis. We emphasize that, although we are applying our fitting technique only to a limited set of observations of RX~J1347.5--1145, it is flexible enough it could be easily extended to other objects and SZ observations, as well as to non-SZ instruments.

The choice of the parametric technique is motivated by the requirement of having a general, self-consistent description of the thermal SZ effect signal independent of the specific type of data considered for the analysis. Indeed, any parametric model can be flexibly adapted to describe both image- and Fourier-space data once the respective set of parameters is defined. Moreover, it offers the possibility of readily deriving quantities relevant for the study of cluster physics and cosmology (e.g. total mass, integrated flux). On the other hand, although computationally demanding, Bayesian inference provides a robust and powerful approach to parametric model reconstruction. For a comprehensive discussion of Bayesian statistics and modelling, see, e.g., \citet{Trotta2008} and references therein.

We here discuss a few crucial aspects of the joint analysis of single-dish and interferometric data. A more extensive presentation of our fitting technique can be found in Appendix~\ref{app:analysis}.

\subsection{Computing the joint likelihood}\label{ssec:jointlike}
One of the crucial steps in a joint Bayesian analysis of multiple observations is the computation of their joint likelihood. In the case of completely independent measurements, it would be enough to consider the product of the likelihood functions of the individual observations. However, potential contamination from astrophysical components other than (and uncorrelated with) the SZ signal explicitly modelled in the analysis could in fact introduce non-negligible covariance between different data sets, and should be accounted for in terms of additional contributions to the generalized noise covariance matrix. This is the case of primary CMB anisotropies, or unresolved sources below the confusion limit of the instrument (see Appendix \ref{ssapp:other} for a discussion). As a consequence, the joint likelihood function in the simplified form is valid for independent data, and calculating it for a mixture of imaging and visibility data may be a non-trivial exercise. Nevertheless, it turns out that for the data employed in our study, the impact of the cross-data correlations on the parameter reconstruction is minimal, and the simplified form of the joint likelihood function consisting of the product of single likelihoods can still be used.

Since the {\it Planck} Compton parameter $\vary$ map has been explicitly built to minimize the signal from astrophysical components other than the SZ effect, we can assume the generalized noise covariance matrix does not include any terms arising from the correlation of the {\it Planck} map with the ACA+ALMA and Bolocam observations. In particular, as reported in \citet{Remazeilles2011}, the high signal-to-noise ratio of the CMB signal in all of the {\it Planck} HFI maps guarantees the MILCA algorithm is able to efficiently remove the corresponding contamination from the recovered thermal SZ map. On the other hand, ACA, ALMA, and Bolocam may in turn not be independent, as any contaminating signal would be common to all the corresponding data sets.  However, as discussed before, the CMB plays a negligible role in the noise budget of the ACA and ALMA measurements when compared to the instrumental noise. Therefore, we assume the contribution to the joint likelihood from off-diagonal blocks of the CMB component of the generalized covariance matrix to be negligible. Furthermore, as in, for example, \citet{Feroz2009}, we assume the confusion from unresolved sources to be characterized by an uncorrelated angular power spectrum.  Hence, considering the limited overlap in the scales probed by the different observations, the confusion covariance matrix can be considered to be block diagonal. The ACA, ALMA, and Bolocam data can then be considered to be independent of each other, and we compute the joint likelihood function as the product of the likelihoods of each data set.

We tested the validity of the above assumptions by running our model reconstruction technique on a sample of mock observations, including either correlated or independent CMB realizations for each of the simulated data sets. In both the cases, we have been able to recover the input model parameters. Moreover, we found no significant difference between the correlated and uncorrelated CMB simulations, therefore allowing each data set to be treated as independent from the others. The joint log-likelihood is therefore computed as the product of the individual likelihoods of each of the data sets presented in the previous sections.

\subsection{Hyperparameters}\label{ssec:hyperpar}
The reconstruction of a model from the simultaneous analysis of multiple observations relies on the assumption of having perfectly calibrated data. However, systematics in the overall calibration may introduce non-negligible relative scaling factors between the different measurements. We can account for possible miscalibration offsets simply by multiplying the models for each of the data subsets by a scaling hyperparameter $\kappa$. 

On the other hand, the statistical uncertainties associated with the image- and Fourier-space observations may suffer from distinct systematic effects that could bias the reconstruction of the model parameters. Therefore, we can weight the likelihood of each of the data sets by a hyperparameter $\eta$, whose estimate is driven directly by the statistical properties of the measurements. As discussed in \citet{Hobson2002a}, such parameter is set to have an exponential prior, as derived by assuming to have no prior knowledge about the weighting factors, apart from the requirement of a unitary expectation value.

\subsection{Implementation details}\label{ssec:implementation}
The fitting pipeline is written in \textsc{python} and uses primarily standard packages (e.g. \textsc{NumPy}, \textsc{SciPy}). The Fourier transforms are computed using the \textsc{FFTW} library \citep{Frigo2012} and its \textsc{python} wrapper \textsc{pyFFTW}. All the common astronomical tasks are managed exploiting the community-developed package \textsc{AstroPy} \citep{Astropy2018}. 

A Monte Carlo Markov Chain (MCMC) approach is adopted for performing the simultaneous forward-modelling of images and visibilities. 
In particular, we make use of the specific implementation of the affine-invariant ensemble sampling technique \citep{Goodman2010} provided by the \textsc{emcee} \textsc{python} package \citep{Foreman2013}. The Message Passing Interface (MPI) protocol is used to parallelize our pipeline.

We evaluate the synthetic visibilities for the extended components using the modelling tool provided in the \textsc{galario} Python library \citep{Tazzari2018}. It can generate an interferometric model by sampling Fourier-space data defined on a regular grid at the positions of the sparse, observed visibilities using a bilinear interpolation algorithm. We modified the core \textsc{galario} library to allow for a more robust and accurate description of sources that extend significantly over the field of view or are at large offsets from the phase centre direction of the interferometric data.

\section{Results}\label{sec:results}
\subsection{Reconstruction of the pressure profile}\label{ssec:results}
We discuss here the results of the reconstruction of a model for the pressure profile of RX~J1347.5--1145 by applying our modelling technique to a set of interferometric and image-domain data of the cluster SZ signal. The ability to discriminate between global thermodynamic properties and local perturbations to them is useful for providing a better understanding of the physical and dynamical state of the cluster. The wide range of spatial scales probed by a combination of single-dish and interferometric observations provides the unique opportunity to build an inclusive description of the physical and thermodynamic state of RX~J1347.5--1145.

RX~J1347.5--1145 is a clear example of a disturbed cluster for which it is not possible to obtain an unambiguous definition of a geometric centre for the ICM distribution.  This is not an isolated case. The optical image is, for example, reminiscent of the Coma cluster, which also possesses two very bright elliptical galaxies separated by some $200~\mathrm{kpc}$ \citep[e.g.][]{Vikhlinin1997}. Unlike RX~1347.5--1145, neither of two galaxies in Coma is embedded in a cool core. As a result, the definition of the cluster centroid in Coma is equally problematic in both X-ray and SZ images, while for RX~1347.5--1145 the centre is often chosen to coincide with the cool core. This issue may be crucial when reconstructing an accurate physical model of a galaxy cluster, since different specific choices of the reference position for modelling the thermodynamic profiles may lead to different implications for the derived cluster properties.

As our baseline pressure model, we employ a generalized Navarro--Frenk--White (gNFW) profile. Hydrodynamic simulations have shown that it can describe reasonably well the radial pressure profile of a galaxy cluster \citep{Nagai2007a}. This motivated its extensive application in a number of SZ studies for parametrizing the observed electron pressure distributions \citep[e.g.][]{Mroczkowski2009,Sayers2013,Sayers2016b,Romero2018,Shitanishi2018}. The gNFW pressure model can be written as a function of radial distance $r$ from the centroid of the galaxy cluster $(x_{\mathrm{g\textsc{nfw}}},y_{\mathrm{g\textsc{nfw}}})$ as follows:
\begin{equation}
    P_{\mathrm{e}}(r) = P_{\mathrm{ei}}\left(\frac{r}{r_{\mathrm{s}}}\right)^{-\gamma}\left[1+\left(\frac{r}{r_{\mathrm{s}}}\right)^{\alpha}\right]^{(\gamma-\beta)/\alpha} \,,
    \label{eq:pressure1}
\end{equation}
where $P_{\mathrm{ei}}$ is a pressure normalization factor and $\gamma$, $\alpha$, and $\beta$ are the slopes at small, intermediate, and large scales with respect to a characteristic radius $r_{\mathrm{s}}$. It is easy to extend the above equation to the case of a cluster with projected eccentricity $\varepsilon$ by substituting the ratio $(r/r_{\mathrm{s}})$ with a generalized ellipsoidal distance $\xi=\xi(r_{\mathrm{s}},\varepsilon)$. The details are presented in Appendix~\ref{app:ellipsoid}.

As discussed at the beginning of this section, RX J1347.5--1145 is thought to have undergone a major merger almost entirely in the plane of the sky. For this reason, we do not include any kinematic SZ contribution to model the cluster sub-components or its bulk motion. We refer to \citet{Zemcov2012}, and \citet{Sayers2016a,Sayers2018} for discussions about the kinematic SZ effect in RX J1347.5--1145.

\subsubsection{Spherically symmetric SZ profile centred at the X-ray peak}\label{sssec:sphere}
The numerous analyses of the X-ray emission from RX~J1347.5--1145 have shown that the peak of the X-ray surface brightness is located at the position of the cool-core region, only a few kiloparsecs away from the wBCG. In the major merger scenario introduced before, this has been generally considered as the centre of the primary cluster component. Therefore, we first apply our fitting pipeline jointly to the ACA+ALMA, Bolocam, and \textit{Planck} data for modelling the thermal SZ signal from the region corresponding to the cool core observed as a strong cusp in the X-ray surface brightness maps of RX~J1347.5--1145. In particular, we fit a spherically symmetric gNW pressure profile whose centroid is set at the position of the peak in the observed X-ray brightness distribution, $(\mathrm{13^h 47^m 30\fs593,-11^d 45^m 10\fs050})$. The parameters are fixed at the values reported in \citet{Arnaud2010} for the sample of a cool-core pressure profile, namely $\alpha=1.2223$, $\beta=5.4905$, and $\gamma=0.7736$. The pressure normalization $P_{\mathrm{ei}}$ and the profile characteristic radius $r_{\mathrm{s}}$ are left free to vary and are assigned wide uninformative uniform priors. Since we are interested in modelling only a portion of the cluster, we do not consider here any prior on the integrated SZ signal. Moreover, the data are not sensitive enough and do not provide sufficient spectral information for deriving any constraint on the electron temperature distribution through the measure of the level of relativistic corrections in the observed SZ effect. Therefore, we fix the electron temperature to $T_{\mathrm{e}}=7~\mathrm{keV}$, characteristic for the inner $\sim 50\;{\rm kpc}$ region around the wBCG as derived from the analysis of {\it Chandra} data (discussed in Section~\ref{sssex:szxray}). In order to remove the contamination from the bright radio source observed at the centre of the cluster, we additionally include in our model a point-like component for which we assume a power-law spectral dependence. All the corresponding parameters --- source coordinates $x_{\mathrm{\textsc{ps}}}$ and $y_{\mathrm{\textsc{ps}}}$, flux normalization $i_{\mathrm{\textsc{ps}i}}$, spectral index $\alpha_{\mathrm{\textsc{ps}}}$ --- are constrained simultaneously with the cluster pressure profile, and set to have uniform priors: the source position is constrained to the most central of the ALMA fields, while we consider a wide uninformative range for the normalization parameter based on previous measurements of the source flux around $100~\mathrm{GHz}$ \citep{Sayers2016a}; similarly, the source spectral index is assumed to be negative but larger than -2, and thus to have a broad prior around the typical value for synchrotron radiation. Furthermore, we include both weighting and scaling hyperparameters in the analysis, providing Gaussian priors for the latter, with unitary mean values and standard deviations equal to the flux calibration uncertainties for each of the data sets.

A summary of the priors on the model parameters can be found in Table \ref{tab:priors}. We find that no biases are introduced by the specific choice of the prior distributions. This has been performed by sampling the posterior distribution obtained in the case of a constant likelihood, as it would be in absence of data. The results of such a data-free run are shown in Fig.~\ref{fig:figure10} in Appendix \ref{app:datafree}.

\begin{table}
    \caption{Summary of the prior distributions for the parameters of the spherical cool-core (Section~\ref{sssec:sphere}) and ellipsoidal (Section~\ref{sssec:ellipse}) models. The parameters $\delta$ and $\mathcal{U}[a,b]$ denote respectively a Dirac delta function and a uniform distribution in the interval $[a,b]$, while $\mathcal{N}[\mu,\sigma]$ is a normal distribution with mean $\mu$ and standard deviation $\sigma$; $\mathcal{E}[1.00]$ is instead the exponential distribution with unitary expectation value discussed in Section~\ref{ssec:hyperpar}. The coordinates of the gNFW and point-source model components are provided in terms of the angular distance from the peak in the X-ray surface brightness, $(\mathrm{13^h 47^m 30\fs593},-\mathrm{11^d 45^m 10\fs050})$, while the flux normalization of the power-law spectral model is measured at the reference frequency $\nu_{\mathrm{\textsc{ps}}}=90~\mathrm{GHz}$. The values of the $\sigma_i$ entering the priors of the scaling hyperparameters $\kappa_i$ for each of the data sets analysed can be found in Section~\ref{sec:data}.}
    \label{tab:priors}
    \begin{center}
    \begin{tabular}{ccrr}
    \hline
    Parameter & Units & Cool-core & Global \\
    \hline
    $x_{\mathrm{g\textsc{nfw}}}$     &             arcsec             &        $\delta[0.00]$        &  $\mathcal{U}[-72.78,72.50]$ \\
    $y_{\mathrm{g\textsc{nfw}}}$     &             arcsec             &        $\delta[0.00]$        &  $\mathcal{U}[-85.04,66.63]$ \\
    $T_{\mathrm{e}}$            &               keV              &        $\delta[7.00]$        &        $\delta[12.50]$        \\
    $P_{\mathrm{ei}}$                &     $\mathrm{keV~cm^{-3}}$     &   $\mathcal{U}[0.00,1.00]$   &   $\mathcal{U}[0.00,1.00]$    \\
    $r_{\mathrm{s}}$                 &              arcmin            &   $\mathcal{U}[0.00,30.00]$  &   $\mathcal{U}[0.00,30.00]$   \\
    $\varepsilon$                    &               --               &          $\delta[0.00]$      &   $\mathcal{U}[0.00,1.00]$    \\
    $\theta$                         &            degrees             &          $\delta[0.00]$      &  $\mathcal{U}[-90.00,90.00]$  \\
    $\alpha$                         &               --               &        $\delta[1.2223]$      &       $\delta[1.2223]$        \\
    $\beta$                          &               --               &        $\delta[5.4905]$      &       $\delta[5.4905]$        \\
    $\gamma$                         &               --               &        $\delta[0.7736]$      &  $\mathcal{U}[0.00,5.00]$     \\
    $x_{\mathrm{\textsc{ps}}}$       &             arcsec             &  $\mathcal{U}[-31.53,30.47]$ &  $\mathcal{U}[-31.53,30.47]$  \\
    $y_{\mathrm{\textsc{ps}}}$       &             arcsec             &  $\mathcal{U}[-40.58,21.42]$ &  $\mathcal{U}[-40.58,21.42]$  \\
    $i_{\mathrm{\textsc{ps}i}}$      &               mJy              &  $\mathcal{U}[0.01,20.00]$  &  $\mathcal{U}[0.01,20.00]$   \\
    $\alpha_{\mathrm{\textsc{ps}}}$  &               --               &   $\mathcal{U}[-2.00,0.00]$  &   $\mathcal{U}[-2.00,0.00]$   \\
    \hline
    $\kappa_i$                       &               --               & $\mathcal{N}[1.00,\sigma_i]$ & $\mathcal{N}[1.00,\sigma_i]$ \\
    $\eta_i$                         &               --               &     $\mathcal{E}[1.00]$      &      $\mathcal{E}[1.00]$      \\
    \hline
    \end{tabular}
    \end{center}
\end{table}

\begin{figure}
    \begin{center}
    \includegraphics[width=\columnwidth]{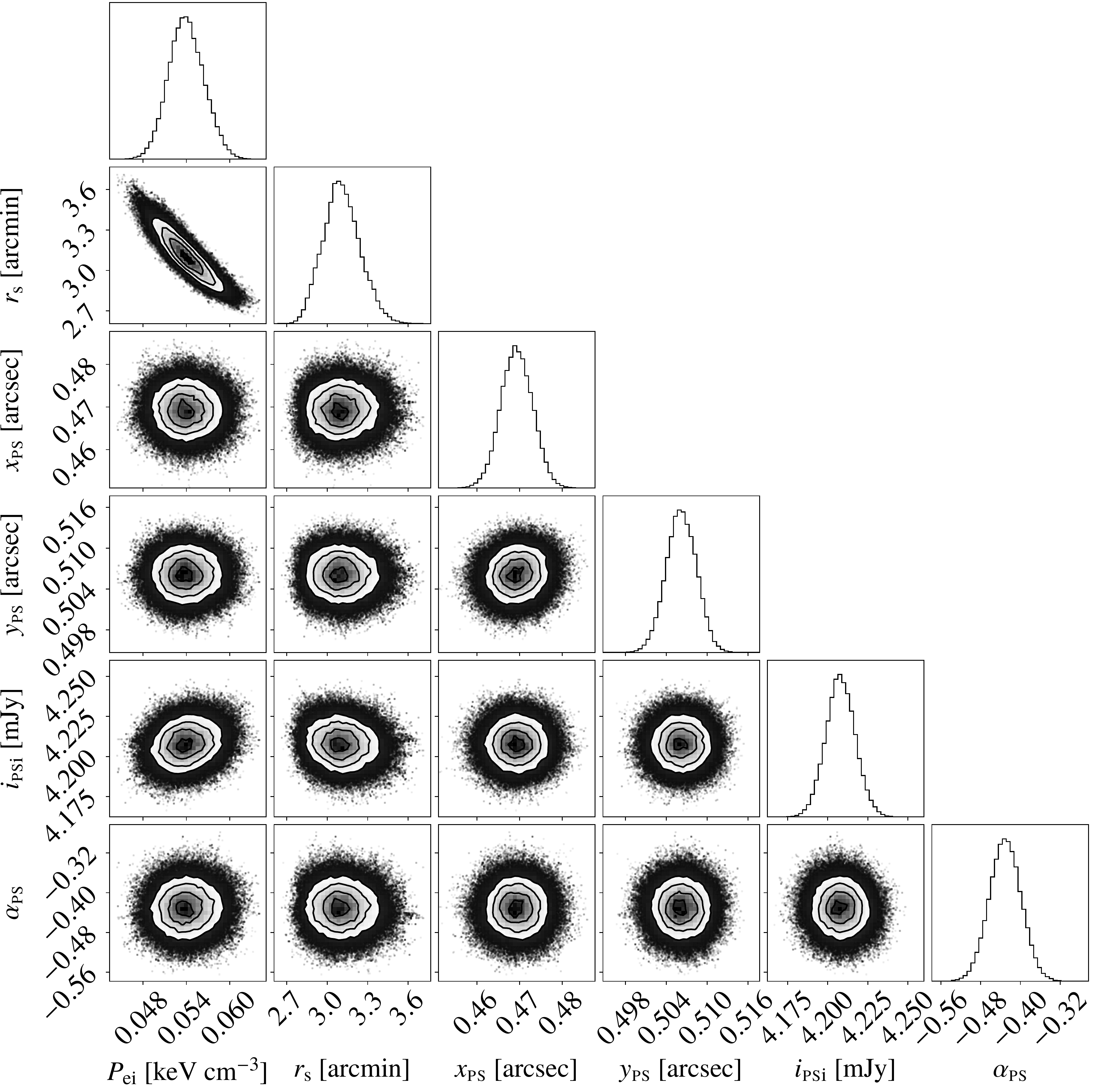}
    \end{center}
    \caption{Bivariate posterior density functions and marginalized distributions of the MCMC parameters from the fit of the spherical gNFW pressure profile ($P_{\mathrm{ei}}$, $r_{\mathrm{s}}$) and power-law point source ($x_{\mathrm{\textsc{ps}}}$, $y_{\mathrm{\textsc{ps}}}$, $i_{\mathrm{\textsc{ps}i}}$, $\alpha_{\mathrm{\textsc{ps}}}$). The phase space was sampled by 200 walkers in 2000 steps after a preliminary burn-in phase of 1000 steps. The reported contours correspond to 68\%, 95\%, and 99\% confidence levels. The values of the respective best-fitting model parameters are presented in Table~\ref{tab:bestfit1}.
    The inferred position and the spectral index of the point-source model are reported as in Table \ref{tab:priors}.}
    \label{fig:figure04}
\end{figure}

\begin{table}
    \caption{Best-fitting parameters and 68\% confidence interval of the spherical gNFW profile and the compact radio source with power-law spectral index. The reference values for the spatial coordinates and the compact source flux are defined as in  Table \ref{tab:priors}.
    }
    \label{tab:bestfit1}
    \begin{center}
    \begin{tabular}{ccrrr}
    \hline
    Parameter   & Units & Mean & 16\textsuperscript{th} perc. & 84\textsuperscript{th} perc.\\
    \hline
    $P_{\mathrm{ei}}$                       & $10^{-2}~\mathrm{keV~cm^{-3}}$ &   $5.39$ &    $5.16$ &    $5.65$ \\
    $r_{\mathrm{s}}$                        &              arcmin            &   $3.10$ &    $2.97$ &    $3.24$ \\
    $x_{\mathrm{\textsc{ps}}}$              &    $10^{-2}~\mathrm{arcsec}$   &  $46.93$ &   $46.57$ &   $47.31$ \\
    $y_{\mathrm{\textsc{ps}}}$              &    $10^{-2}~\mathrm{arcsec}$   &  $50.62$ &   $50.40$ &   $50.84$ \\
    $i_{\mathrm{\textsc{ps}i}}$             &               mJy              &  $4.208$ &   $4.199$ &   $4.217$ \\
    $\alpha_{\mathrm{\textsc{ps}}}$         &               --               & $-0.431$ &  $-0.462$ &  $-0.399$ \\
    \hline
    \end{tabular}
    \end{center}
\end{table}

\begin{figure}
  \includegraphics[width=\columnwidth]{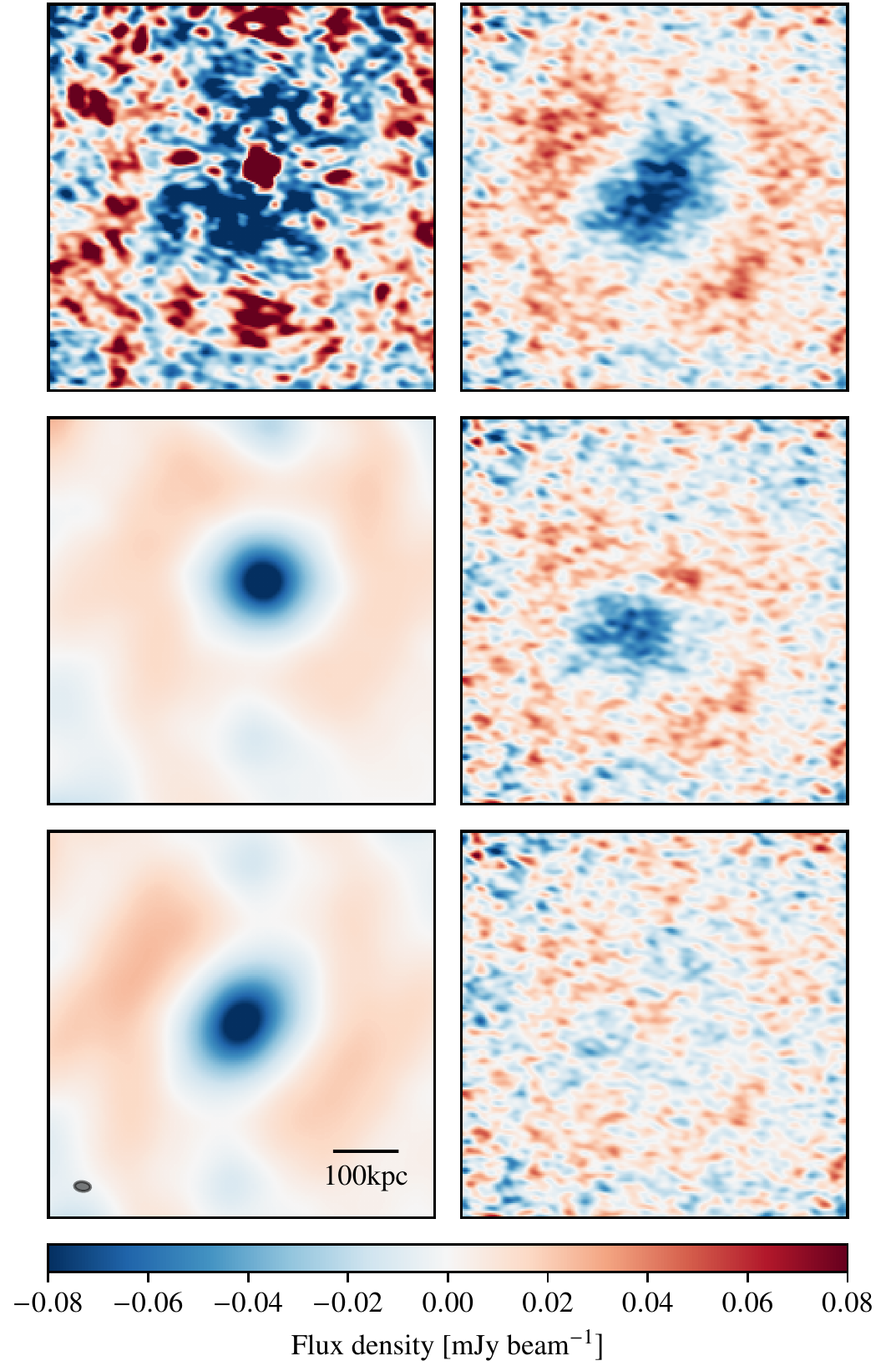} 
  \caption{Images from the raw (top left) and point-source subtracted ACA+ALMA data (top right), X-ray centred spherical (middle left) and free-centroid ellipsoidal (bottom left) gNFW models and corresponding residual maps (middle and bottom right, respectively). All the images are the dirty maps of the corresponding visibilities, and are generated adopting a multi-frequency mosaic gridding approach with natural weighting of all the fields and spectral windows from both ACA and ALMA data. The resulting synthesized beam ($4.11~\mathrm{arcsec} \times 2.44~\mathrm{arcsec}$ at a position angle of $83.4\degr$) and the reference spatial scale for the six maps are reported in the bottom left panel.}
  \label{fig:figure05}
\end{figure}

The posterior probability density function resulting from the MCMC sampling of the parameter space is shown in Fig.~\ref{fig:figure04}. The best-fitting parameters and the corresponding uncertainties for both model components are defined by considering the medians and the central credibility intervals of the marginalized posterior distribution of each model parameter (see Table~\ref{tab:bestfit1} for a summary). A synthetic realization for each of the observations employed in the modelling process is then generated and subtracted from the raw data sets. The poorer resolution and sensitivity of both the {\it Planck} and Bolocam data limit the possibility of observing any significant residual structure. Therefore, although the analysis has been performed jointly on all the available SZ data sets, we present in Fig.~\ref{fig:figure05} only the maps obtained from the residual ACA+ALMA measurements. Notice that the model subtraction from interferometric data is performed directly in visibility space, and we only show the dirty images of the processed measurements for illustrative purposes. 

The map obtained from the model-subtracted visibilities (middle right-hand panel of Fig.~\ref{fig:figure05}) clearly shows a residual SZ signal at the position of the south-eastern substructure. Consistently with previous analyses \citep{Adam2014,Ueda2018}, the reconstruction of the spherically symmetric SZ model centred at the X-ray peak leads to the conclusion that the X-ray and SZ maps have coincident excesses, suggesting that this region is strongly overpressurized with respect to the cluster core. 

Conversely, it is not possible to observe any signature of strong local deviations from the smooth gas pressure distribution within the cool core of the cluster. Indeed, no excess in the model-subtracted ACA+ALMA maps of the SZ signal is significantly detected in the direction of the region around the wBCG. In agreement with \citet{Ueda2018}, this suggests the occurrence of subsonic sloshing motions within the cool core.

\subsubsection{Comparison of the spherically symmetric SZ and X-ray profiles centred at the X-ray peak}\label{sssex:szxray}
\begin{figure}
    \begin{center}
    \includegraphics[width=\columnwidth]{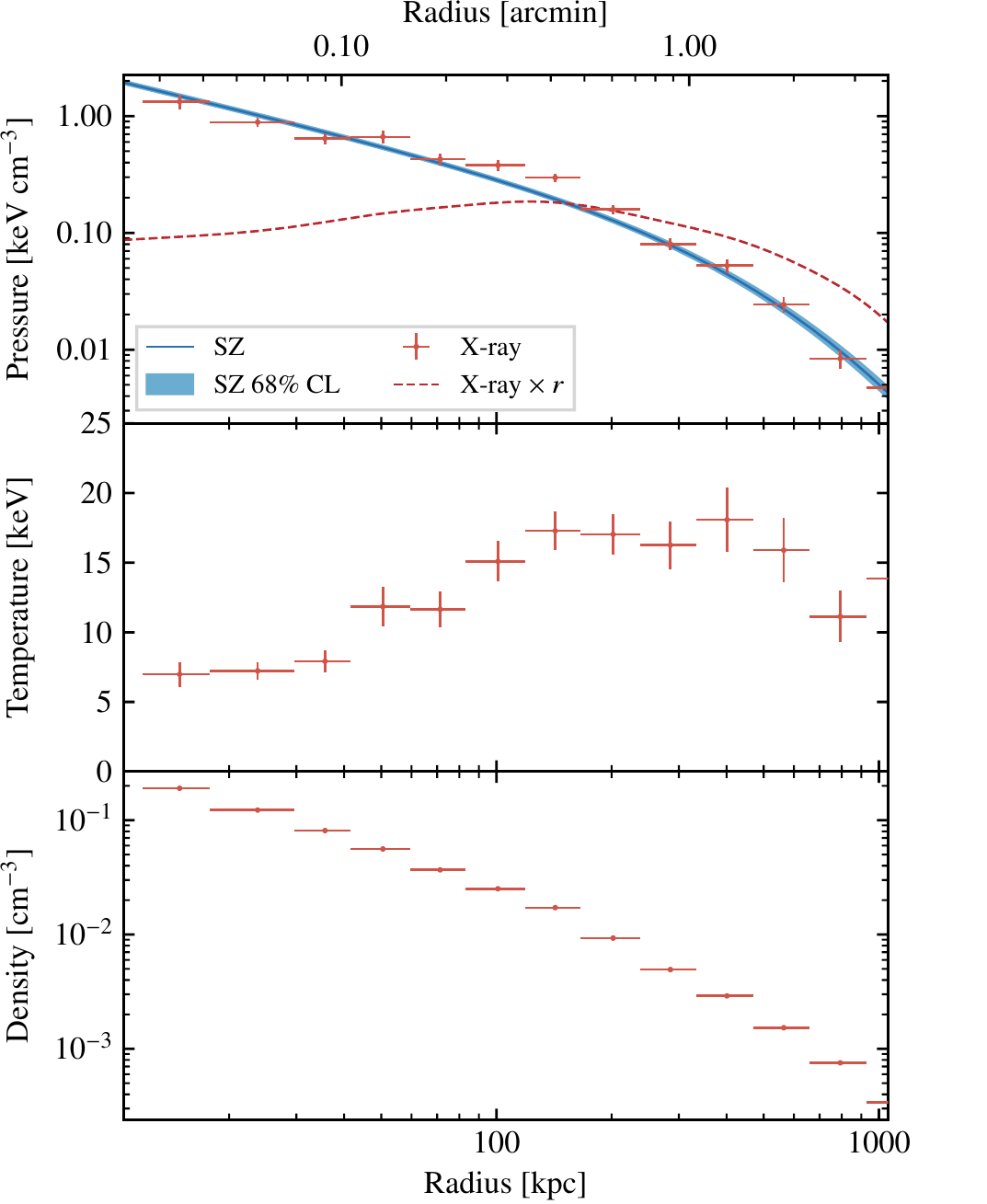}
    \end{center}
    \caption{Azimuthally-averaged radial profiles of the deprojected thermodynamic properties of RX~J1347.5--1145 based on {\it Chandra} X-ray data. The red, dashed line in the top panel represents the product of the radial distance and the pressure profile obtained by interpolating the deprojected pressure radial distribution. For a comparison, we also report the pressure model (blue line in the top panel) derived through the joint image-visibility analysis of the SZ observations discussed in Section~\ref{sssec:sphere}.}
    \label{fig:figure06} 
\end{figure}

In this section, we compare the radial pressure profiles inferred through the joint image-visibility reconstruction of the spherical gNFW model and from the independent analysis of {\it Chandra} X-ray data. We employ archival Chandra observations (OBSIDs: 3592,13516,13999,14407). The 0.5-3.5 keV image is shown in the middle panel of Fig.~\ref{fig:figure01}. We obtain from the X-ray data an estimate of the radial profile of the cluster electron pressure by multiplying the radial profiles of deprojected electron density and temperature distributions generated using the procedure described in \citet{Churazov2003}.
A plot of the thermodynamic profiles of RX~J1347.5--1145 is shown in Fig.~\ref{fig:figure06}. The pressure model determined using the SZ observations is overlaid in the top panel, and shows good overall agreement with the X-ray profile within the uncertainties of the two independent determinations. However, an excess in the pressure distribution derived from the X-ray data with respect to the SZ model can be seen at $\sim30~\mathrm{arcsec}$ from the centre, roughly at the distance where the south-eastern substructure is located. Such modest discrepancy is not surprising, given that any departures from the spherical symmetry may affect differently the radial pressure profiles derived from X-ray and SZ data. Note, also, that even the relativistic correction alone could modify the normalization of the derived Compton parameter $\vary$ by around $6\%$, if the change of the temperature from $7~\mathrm{keV}$ in the core to $20~\mathrm{keV}$ some $20-30~\mathrm{arcsec}$ away was properly taken into account.

The broad agreement between the radial pressure profiles from the independent X-ray and SZ analyses provides a partial validation of our joint image-visibility modelling. It also suggests that the perturbations present in the cluster gas are not so extreme as to affect dramatically the reconstruction of the pressure profile at all radii \citep[see e.g.][for a discussion of biases arising from inhomogeneities in the gas within galaxy clusters]{Khedekar2013}.

\subsubsection{SZ-driven ellipsoidal model with free centroid}\label{sssec:ellipse}
From the image of the X-ray surface brightness in Fig.~\ref{fig:figure01}, it is evident that the peak of the X-ray emission is unambiguously associated with the cool and dense region around wBCG. Also, the pressure profile derived from the X-ray data (see Fig.~\ref{fig:figure06}) clearly shows that the gas pressure is increasing towards the X-ray peak. The question arises whether we should expect a very prominent peak in the SZ signal at exactly the same location, which will then dominate the overall SZ signal. To answer this question we plot in Fig.~\ref{fig:figure06} the interpolated X-ray pressure profile $P(r)$ multiplied by $r$ (dashed line). This quantity, $P(r)\times r$, characterizes the contribution of a region with size $\sim r$ to the projected pressure map, i.e., the amplitude of a peak in the SZ images. It is found to be a growing function of the radius up to $r\sim 100\;{\rm kpc}$, implying that the central region is playing a sub-dominant role when compared to scales of the order of few hundreds of kiloparsecs in the projected pressure map. Considering that the typical inner slope for the pressure profile in cool-core clusters $\gamma$ is less than 1, it is not surprising that no strong cusp is expected at the position of the cool core.

The rather modest contribution of the core gas to the overall SZ effect signal makes the definition of a cluster centre from the sole inspection of RX J1347.5--1145 SZ images ambiguous. As seen when imaging the point source-subtracted visibilities in Fig.~\ref{fig:figure05}, the distribution of the SZ signal is indeed fairly smooth across the cool-core region around wBCG. In fact, recent works generally agree that the SZ signal peaks at a location offset south-east of the X-ray surface brightness peak \citep[e.g.][]{Kitayama2016}. We therefore relax all priors on the centroid position and the assumption of spherical symmetry and try to build a model for the pressure distribution based solely on the SZ data. We use the modelling set-up adopted in Section~\ref{sssec:sphere}, but we substitute the spherical gNFW distribution, with centre fixed to the X-ray peak, with a free-centroid elliptical gNFW pressure profile, allowing for eccentricity and arbitrary orientation on the plane of the sky. The coordinates of SZ centroid are bounded to the combined ACA+ALMA mosaicked field of view by the introduction of uniform priors. Along with the pressure normalization $P_{\mathrm{ei}}$ and the profile characteristic radius $r_{\mathrm{s}}$, we further allow the inner slope of the gNFW profile $\gamma$ to vary. Again, the other two indices are fixed to the cool-core values of \citet{Arnaud2010}. For this analysis, we now assume an electron temperature of $12.5~\mathrm{keV}$, estimated by averaging the X-ray temperature profile of Fig.~\ref{fig:figure06} within $1~\mathrm{arcmin}$ from the position of the X-ray peak. Finally, we fit for the cylindrically integrated Compton $\vary$ by assigning a Gaussian prior based on the value derived from the {\it Planck} MILCA Compton $\vary$ map. All the other free parameters of the gNFW model are assigned wide uninformative uniform priors. The specific details of the above priors on the model parameters are listed in Table \ref{tab:priors}. Again, the prior-only sampling shows no biases in the reconstruction of the model parameters due to the assumption on the corresponding prior distributions (see Fig.~\ref{fig:figure11} in Appendix \ref{app:datafree}).

\begin{figure*}
    \begin{tikzpicture}
    \node(a){
        \includegraphics[trim=0 0.50cm 0 0,clip,width=\textwidth]{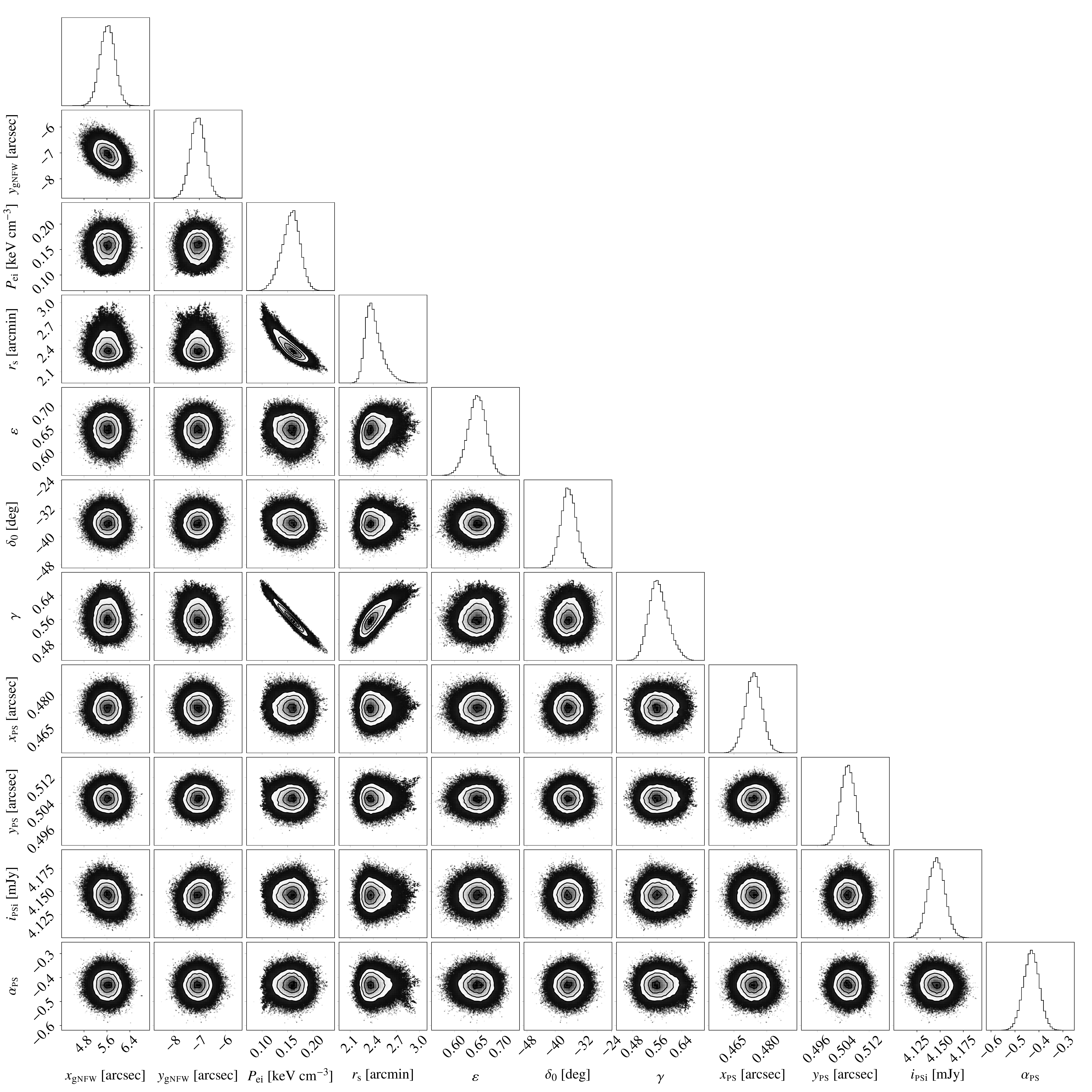}
    };
    \node at (a.north east)
    [anchor=north east,xshift=0mm,yshift=0mm]{
            \begin{tabular}{crrrr}
            \hline
            Parameter   & Units & Mean & 16\textsuperscript{th} perc. & 84\textsuperscript{th} perc.\\
            \hline
            $x_{\mathrm{g\textsc{nfw}}}$    &        $\mathrm{arcsec}$       &    $5.62$ &    $5.39$ &    $5.87$ \\
            $y_{\mathrm{g\textsc{nfw}}}$    &        $\mathrm{arcsec}$       &   $-7.06$ &   $-7.34$ &   $-6.77$ \\
            $P_0$                           & $10^{-1}~\mathrm{keV cm^{-3}}$ &    $1.57$ &    $1.38$ &    $1.74$ \\
            $r_{\mathrm{s}}$                &              arcmin            &    $2.39$ &    $2.31$ &    $2.52$ \\
            $\varepsilon$                   &               --               &   $0.648$ &   $0.628$ &   $0.667$ \\
            $\theta$                        &             degree             &   $-36.2$ &   $-38.4$ &   $-34.1$ \\
            $\gamma$                        &               --               &   $0.563$ &   $0.534$ &   $0.598$ \\
            $x_{\mathrm{\textsc{ps}}}$      &    $10^{-2}~\mathrm{arcsec}$   &   $47.40$ &   $47.02$ &   $47.78$ \\
            $y_{\mathrm{\textsc{ps}}}$      &    $10^{-2}~\mathrm{arcsec}$   &   $50.55$ &   $50.32$ &   $50.77$ \\
            $i_{\mathrm{\textsc{ps}i}}$     &               mJy              &   $4.146$ &   $4.137$ &   $4.155$ \\
            $\alpha_{\mathrm{\textsc{ps}}}$ &               --               &  $-0.431$ &  $-0.465$ &  $-0.399$ \\
            \hline
            \end{tabular}
    };
    \end{tikzpicture}
    \caption{{\it Main:} same of Fig.~\ref{fig:figure04} but for the case of the ellipsoidal gNFW pressure profile. Due to the larger number of parameters, we had to increase the burn-in and sampling phases to 4000 and 8000 steps, respectively. {\it Inset table:} as in Table~\ref{tab:bestfit1}, the inset table in the top-right corner summarizes the corresponding best-fitting parameters.}
    \label{fig:figure07}
\end{figure*}

As with the previous case of the spherical profile, Fig.~\ref{fig:figure07} shows the posterior probability density function of the sampled parameters, while a summary of the best-fitting model parameters is reported in the table inset in Fig.~\ref{fig:figure07}. The cluster pressure distribution appears to be described by a slightly eccentric profile. The inner slope of the gNFW model is found to be steeper than that reported by \citet{Arnaud2010} for both the universal and morphologically disturbed profiles, but still lower than for the cool-core sample of clusters. We tested this result by varying the intermediate and outer slopes, but found no significant changes in the estimated value of the inner parameter.

The map of the model-subtracted interferometric data, together with the image of the inferred best-fitting SZ distribution, is presented in the bottom panels of Fig.~\ref{fig:figure05}. No residual structures highlighting possible overpressure in the ICM within RX~J1347.5--1145 are detected at a significant level with respect to the image noise. In particular, the residual amplitude of the SZ effect to the south-east is dramatically reduced when shifting the centroid away from the X-ray peak and allowing for ellipticity. This suggests that the SZ excess
may be at least partially ascribed to purely geometric effects, which is a consequence of the intrinsic eccentricity of RX~J1347.5--1145 in the inner $\sim200~\mathrm{kpc}$ region. It is worth noting that the centroid of our ellipsoidal pressure model is consistent with the position of the SZ peak reported by \citet{Kitayama2016}. Since  the presence of strong local overpressures may easily result in a non-negligible offset between these positions, such fair agreement further suggests that the pressure structure may be more regular than could be derived from X-ray analyses.

\subsubsection{Compact radio source}\label{sssec:compact}
We found that the reconstruction of the model for the central radio source is independent of the specific profile used to fit the underlying SZ signal. The position and the spectral index from the spherical and ellipsoidal profile fits are entirely consistent, while the discrepancy between the two estimates of the flux normalization is  within the flux calibration uncertainties of the ACA+ALMA data. Moreover, the constraints on the central radio source are in good agreement with the parameters derived by \citet{Kitayama2016} using the same interferometric observations of this work, although our best-fitting model has a slightly steeper radio spectrum and larger normalization at the same reference frequency of $92~\mathrm{GHz}$. Nevertheless, these differences are not enough to solve the tension with previous studies, which report fluxes of $4.9\pm0.1~\mathrm{mJy}$ at $86~\mathrm{GHz}$ \citep[CARMA;][]{Plagge2013}, $4.4\pm0.3~\mathrm{mJy}$ at $140~\mathrm{GHz}$ and $3.2\pm0.2~\mathrm{mJy}$ at $240~\mathrm{GHz}$ \citep[Diabolo;][]{Pointecouteau2001} against our corresponding estimates of respectively $4.29\pm0.01~\mathrm{mJy}$, $3.48\pm0.05~\mathrm{mJy}$, and $2.76\pm0.09~\mathrm{mJy}$ at $68\%$ confidence level. The determination of the point-source parameters is principally driven by the interferometric data, and we assess that no significant bias is introduced as a consequence of a possible miscalibration of the ACA or ALMA measurements. For this reason, we repeat the above analysis on the ACA, ALMA and {\it Planck} observations, fixing all the model parameters to the best-fitting values of Table~\ref{tab:bestfit1} but without marginalising over the scaling hyperparameters (Sec.~\ref{ssec:hyperpar}). We exclude Bolocam from the test to avoid any systematics related to the unresolved radio source, while deriving a constraint on the scaling parameter mainly based on the cluster SZ signal. Since {\it Planck} data provide more frequency channels than parameters necessary to describe the SZ contribution, we assume any potential radio contamination is marginalized over as a result of the MILCA component separation. Further, the radio source flux is less than 2\% of the total SZ flux on $15~\mathrm{arcmin}$ scales (Eq.~\ref{eq:planck1}), and residual radio source contamination is well within the statistical uncertainty of the {\it Planck} measurement.
The modelling provides estimates of the {\it Planck} and ALMA+ACA scaling hyperparameters with ratio equal to $0.993\pm 0.038$, therefore consistent with unity and supporting the general scenario that ascribes the discrepancies in the flux measurements to a long-term variability of the radio source \citep{Kitayama2016}. However, we have not been able to characterize any possible time dependencies due to the limited sampling over time provided by the observations used in our analysis.

\subsection{Interpretation and discussion}\label{ssec:xclues}
The joint image-visibility SZ analysis has shown that it is possible to account for the SZ signal from the south-eastern SZ excess simply by allowing the centroid to vary freely, away from the X-ray peak, and by adopting an ellipsoidal model to describe the cluster pressure profile. This would imply that pressure distribution is more regular than one would derive when treating the south-eastern excess seen in the X-ray images as a significant overpressure with respect to the cluster cool core. While a fraction of the SZ effect from the south-eastern structure can certainly be ascribed to the elongated morphology of RX~J1347.5--1145, it is important to consider that any signatures of overpressurized gas could possibly be pushed below the image noise level as a consequence of the overfitting of the SZ signal. This may indeed be a crucial issue arising due to the larger number of free parameters adopted when considering the free-centroid ellipsoidal model instead of the spherically symmetric pressure profile. It is worth highlighting that the two models are meant to describe different physical components, and a direct comparison based on statistical considerations would provide misleading results. To do this properly, the inclusion with the spherical model of an additional component to describe the south-eastern substructure would be required. However, this would likely not provide a good description of RX~J1347.5--1145, given the complex morphology of this merging cluster. Nevertheless, the lack of significant residuals in the free-centroid model-subtracted interferometric map represents an interesting result. The possibility of describing the pressure substructure observed south-east of the X-ray peak simply by means of a different model geometry implies the ICM in RX~J1347.5--1145 may be closer overall to pressure continuity than has been discussed in the previous studies. Specifically, this could be interpreted as hinting at a less violent merger history, or it could indicate that the disturbed X-ray morphology is a result of the merger being in a late stage. However, the SZ data employed in our analysis are not able to entirely rule out any of the above scenarios. Thus, we consider this result as further motivation for the interpretation we propose.

In order to gain a more thorough understanding of the physical properties of RX~J1347.5--1145, we then compare the results from the SZ study to what can be inferred from {\it Chandra} X-ray imaging and spectroscopic analysis. Since the resolution of both Bolocam and {\it Planck} maps is too poor to allow for a direct comparison with X-ray data, we hereafter consider only the images generated using the ACA+ALMA data.

\begin{figure*}
    \begin{center}
    \includegraphics[width=\textwidth]{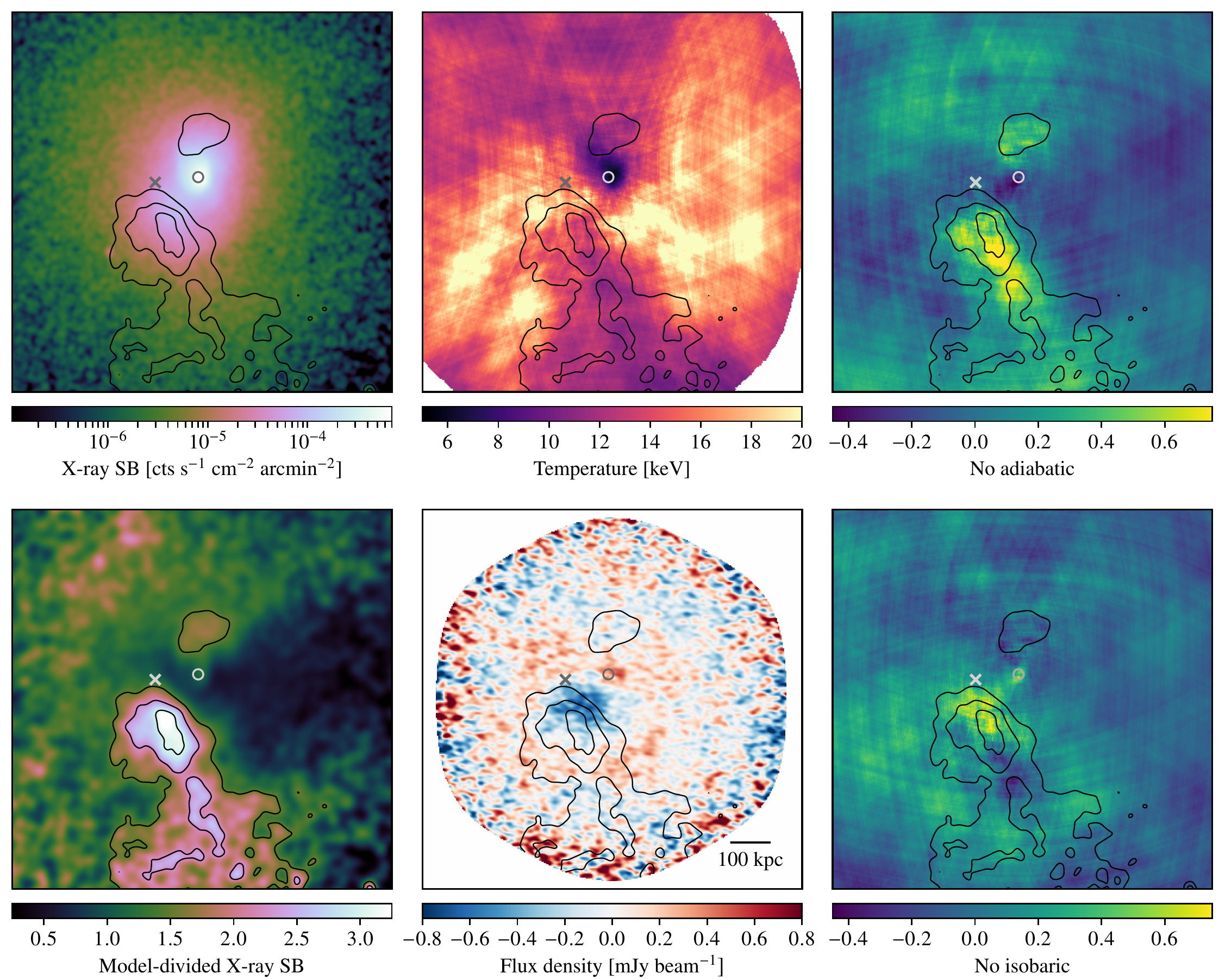}
    \end{center}
    \caption{Raw (top left) and spherical model-divided (bottom left) X-ray surface brightness, temperature (top middle), spherical model-subtracted SZ effect (bottom middle), and X-ray images without adiabatic and isobaric perturbations (top and bottom right, respectively). The solid lines mark the significant structures in the model-divided X-ray map. As in Fig.~\ref{fig:figure01}, the circle and the cross denote the positions of the brightest cluster galaxies wBCG and eBCG, respectively. We report in the bottom-middle panel the reference scale for all the above maps.}
    \label{fig:figure08}
\end{figure*}

\subsubsection{X-ray imaging constraints on the nature of the ICM perturbations}\label{sssec:arithmetic}
The `X-ray arithmetic' method reported in \cite{Churazov2016} allows for the determination of the properties of small perturbations in X-ray images and, in particular, to differentiate between predominantly adiabatic and predominantly isobaric perturbations \citep[see also][]{Arevalo2016,Zhuravleva2016}. This approach uses X-ray images in two different energy bands (typically 0.5-3.5~keV and 3.5-7.5~keV for {\it Chandra} data) and identifies in each image deviations relative to a suitable smooth underlying model. For adiabatically compressed regions, the fluctuations in density and temperature are correlated, while they are anti-correlated in the case of isobaric structures. As a consequence, the perturbations have different amplitudes in the two images, since the emissivity in the harder band is more sensitive to temperature variations. The relation between the amplitudes in two energy bands can be easily predicted for different types of perturbations, and it is straightforward to make a linear combination of two images that completely suppresses the perturbations of one type, leaving the amplitude of the other type unchanged. Note that the perturbations are identified in the maps divided by smooth underlying models, and therefore the prominence of a perturbation will depend on the accuracy of the model choice \citep[for a similar case where model choice significantly affects the level of derived perturbations, see][]{Bonafede2018}.

The application of X-ray arithmetic to RX~J1347.5--1145 is illustrated in Fig.~\ref{fig:figure08}. The top right-hand panel shows the map from which adiabatic perturbations have been removed, revealing a prominent elongated structure to the south-west of eBCG.  This coincides with the most prominent asymmetric excess seen in the X-ray image and, more clearly, in that divided by a spherically symmetric model centred at the X-ray peak (see top left- and bottom left-hand panels in the same figure). When calculating the best-fitting symmetric model, a $90\degr$ wedge to the south-east from wBCG was excluded from the analysis. The isobaric nature of this excess confirms the interpretation of this structure as low-entropy gas stripped from the subhalo associated with the eBCG and embedded in higher entropy ambient gas. Yet another isobaric structure is seen to the north of the wBCG, which could be due to subsonic sloshing of the gas.

The bottom-right panel shows the map free from isobaric perturbations. The remaining structures are less prominent than the isobaric ones. The most prominent adiabatically compressed region is located just in front of the stripped gas, halfway to the position of the eBCG. A comparison of these two images (``no adiabatic'' and ``no isobaric'') suggests that the stripped gas forms an almost isobaric tail, but is moving with a substantial velocity to produce the overpressurized region ahead of it, which can also be identified in the temperature map. Thus, we do not expect the isobaric tail to produce any strong signal in the SZ map. The adiabatic region should instead show up as a local increase of the SZ signal. This is consistent with the presence of the south-eastern excess in the model-subtracted SZ image (middle right-hand panel of Fig.~\ref{fig:figure05}) when considering a spherically symmetric pressure model. In fact, the spatial correlation of the adiabatically compressed gas with the SZ structure is especially convincing when directly comparing the residual SZ substructures with  isobaric-free X-ray maps  (see Fig.~\ref{fig:figure08}). However, the resulting residual is seen to be more extended than the adiabatically compressed region shown in the bottom right-hand panel of Fig.~\ref{fig:figure08}, and to be slightly shifted towards the wBCG. We note, however, that this may arise as a consequence of the different line-of-sight dependencies of the X-ray and the SZ effect, which cause the latter to be generally observed in regions wider than X-ray emissions. On the other hand, the lack of any significant excess after subtracting the best-fitting ellipsoidal SZ model (bottom right-hand panel of Fig.~\ref{fig:figure05}) supports the result that the south-eastern substructure is likely dominated by isobaric rather than adiabatic perturbations. 

To make a crude estimate of the SZ signal expected from the south-eastern excess, we assume that it originates from a sphere of overpressurized gas with radius $\sim 90~\mathrm{kpc}$, shifted away from the wBCG by $\sim 130~\mathrm{kpc}$ in the plane of the sky. Integrating the model of the X-ray emission based on the thermodynamic profiles shown in Fig.~\ref{fig:figure06}, we infer that an excess $\delta I_{\mbox{\tiny X}}/I_{\mbox{\tiny X}} \sim 2$ observed in the X-ray surface brightness (see bottom left-hand panel of Fig.~\ref{fig:figure08}) requires a density perturbation $\delta \rho/\rho\sim 1$. If we assume that the perturbation is fully adiabatic, the pressure in the sphere would then be increased by a factor $ \left[(\rho+\delta \rho)/\rho \right ]^{5/3}\approx 3$. The corresponding enhancement of the SZ signal (integrated pressure profile along the line of sight in the direction of the sphere) is then $\delta \vary/\vary\sim 0.8$. If, instead, we use the X-ray surface brightness excess $\delta I_{\mbox{\tiny X}}/I_{\mbox{\tiny X}}\sim 0.7$ seen in the ``adiabatic'' image (bottom right-hand panel in Fig.~\ref{fig:figure08}), the expected SZ excess is $\delta \vary/\vary\sim 0.3$. These are of course only an order-of-magnitude estimates, given the complexity of the cluster and the assumptions made. Nevertheless, it can be compared to the value of $\delta \vary/\vary\sim 0.24$ obtained by dividing the ACA+ALMA residual map in the middle right-hand panel of Fig.~\ref{fig:figure05} by the spherical gNFW pressure profile and averaging over the circular region corresponding to the model gas sphere introduced before. This would suggest that, when subtracting the spherical model centred at the X-ray peak, it is more likely that the south-eastern substructure seen in the X-ray and SZ maps is predominantly related to isobaric rather than adiabatically compressed gas. Furthermore, this result is consistent with the previous one, that a simple, smooth ellipsoidal pressure profile is sufficient for describing at least partially the observed SZ excess. Anyway, it is worth noting that we do not interpret the nature of the south-eastern SZ structure to be wholly isobaric, but posit that the adiabatic component of the ICM perturbations cannot be solely responsible for the observed structure. The clear spatial coincidence discussed above of the adiabatically compressed gas with the SZ excess would indeed support the possibility of a small contribution from the adiabatic perturbation. Of course, the value of fractional Compton $\vary$ should be treated with caution, since a non-negligible level of contamination from the side lobes and the missing large-scale flux may reduce the actual amplitude of the SZ effect from the excess. Furthermore, the validity of the X-ray arithmetic methodology is in principle limited to small linear perturbations, while, in our study, we are employing it to characterize perturbations in a non-linear regime. As a result, the quantitative estimates of their amplitudes may be inaccurate, although still valid at the order-of-magnitude level. Moreover, we note that the method is able to provide the correct qualitative classification of the adiabatic or isobaric nature of the gas perturbations.

\subsubsection{Gas velocities from X-ray temperatures}
Additional information about the dynamical state of RX~J1347.5--1145 can be derived by analysing its temperature distribution. The morphology of the X-ray temperature map, shown in the top middle panel of Fig.~\ref{fig:figure08}, is reminiscent of the characteristic pattern produced by two subclusters moving with respect to each other with a non-zero impact parameter \citep[see e.g. Fig.~7 in][and Forman et al., in prep.]{Ricker2001}. Indeed, the cooler structures to the north of the wBCG and to the south-west of the eBCG could be associated with the low entropy gas initially bound to infalling subhaloes and now trailing them. For the eBCG, the cooler gas has apparently already been stripped away. On the other hand, the hotter gas is observed to form an ``S''-like pattern between the two subcomponents. 

One can use the measured gas temperatures in order to constrain the velocities of the subhaloes. Indeed, for a body moving steadily through the homogeneous medium\footnote{See \cite{Zhang2019} for the discussion of non-steady motion in a medium with pressure/density gradients} and ignoring for simplicity the contribution of the gravitational potential, the initial gas temperature $T_1$ and the temperature $T_{\mathrm{st}}$ at the stagnation point in front of the body are linked by the Bernoulli equation,
\begin{equation}
\frac{\varv^2}{2}+\frac{\gamma_{\textsc{p}}}{\gamma_{\textsc{p}}-1}\frac{kT_1}{\mu m_{\textsc{p}}}=\frac{\gamma_{\textsc{p}}}{\gamma_{\textsc{p}}-1}\frac{kT_{\mathrm{st}}}{\mu m_{\textsc{p}}},
\end{equation} 
where $\varv$ is the velocity of the body, $\gamma_{\textsc{p}}$ is the polytropic exponent, $\mu\approx 0.61$ is the mean atomic weight, and $m_{\textsc{p}}$ is the proton mass. 
For subsonic motion with respect to the sound speed in the gas with temperature $T_1$, i.e., $c_{\mathrm{s},1}=\sqrt{\gamma_{\textsc{p}}{k_{\mathrm{\textsc{b}}}T_1}/{\mu m_{\textsc{p}}}}$, the temperature gradually increases from $T_1$ far from the body to $T_{\mathrm{st}}$ at the stagnation point. Much of the temperature variation occurs over spatial scales comparable to the size of the body, where the velocity changes significantly. In terms of the temperature ratio, the Bernoulli equation yields
\begin{equation}
    \frac{T_{\mathrm{st}}}{T_1}= 1+\frac{\gamma_{\textsc{p}}-1}{2} \mathcal{M}^2,
    \label{eq:bernoulli}
\end{equation} 
where $\mathcal{M}=\varv/c_{\mathrm{s},1}$ is the body Mach number. 
When the velocity of the body is instead supersonic, a bow shock forms in front of it (see, e.g., \citealt{Keshet2016} or \citealt{Zhang2019} for astrophysical applications). In this case, the gas temperature is $T_1$ ahead of the shock and jumps at the shock front to the temperature $T_{\mathrm{sh}}$, which is related to $T_1$ via the Rankine--Hugoniot condition
\begin{equation}
    \frac{T_{\mathrm{sh}}}{T_1}=\frac{[2\gamma_{\textsc{p}}\mathcal{M}^2-(\gamma_{\textsc{p}}-1)][(\gamma_{\textsc{p}}-1)\mathcal{M}^2+2]}{(\gamma_{\textsc{p}}+1)^2 \mathcal{M}^2}.
    \label{eq:rhcond}
\end{equation}
Between the bow shock and the stagnation point, the temperature increases steadily from $T_{\mathrm{sh}}$ to $T_{\mathrm{st}}$.

\begin{figure}
  \begin{center}
    \includegraphics[width=\columnwidth]{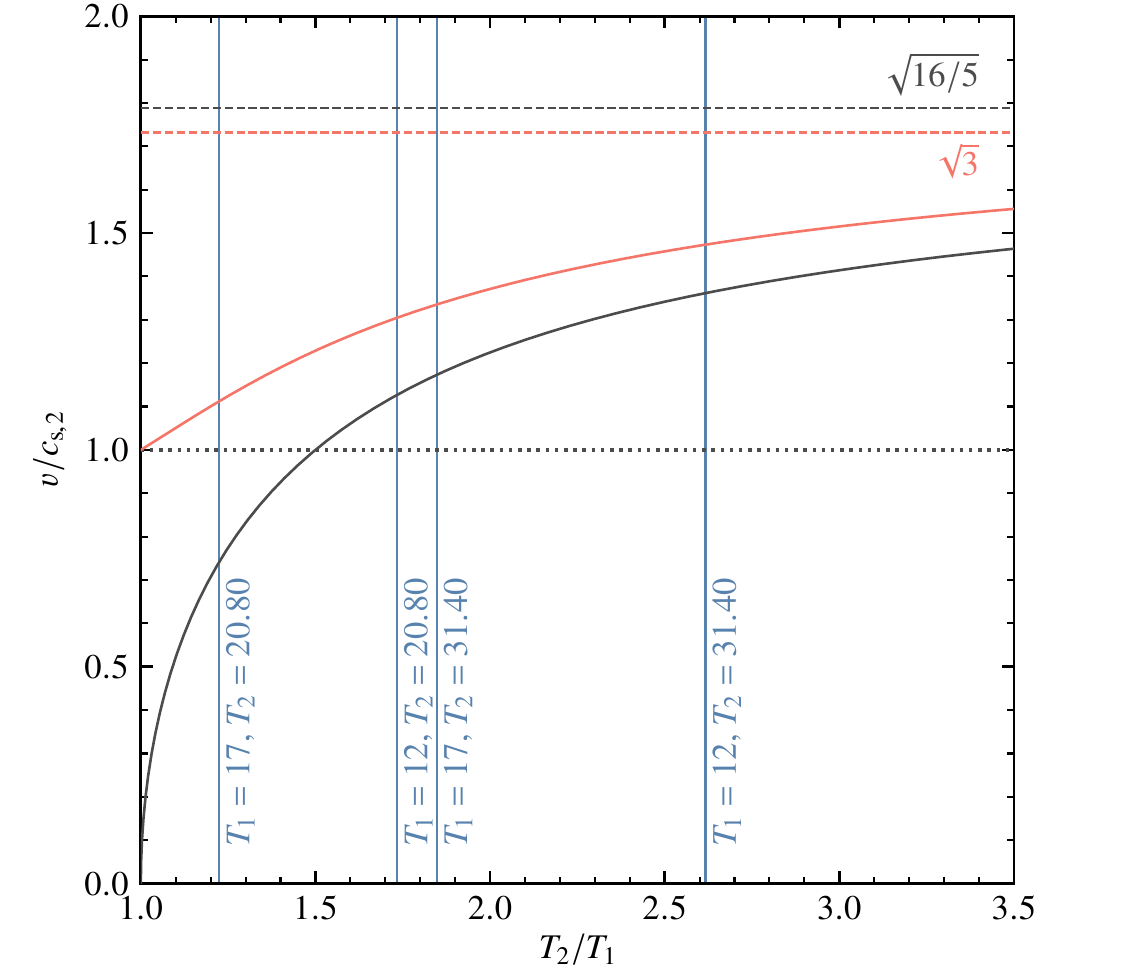}
  \end{center}
  \caption{Relation between the temperature contrast and the body velocity for a steady motion in a homogeneous medium. $T_1$ is the upstream temperature far from the body, while $T_2$ is either the temperature at the stagnation point $T_{\mathrm{st}}$, or at the downstream side of the shock $T_{\mathrm{sh}}$. 
  The body velocity is scaled by the sound speed in the gas with the temperature $T_2=T_{\mathrm{st}}$ (black curve) and $T_2=T_{\mathrm{sh}}$ (red curve). The blue vertical lines show the observational constraints coming from the {\it Suzaku} and {\it Chandra} data. See text for details. The intersections of the blue lines with the black and red lines show the velocity needed to provide the observed temperature ratio.}
  \label{fig:figure09}
\end{figure}

It is then clear that the temperature ratio can be used to infer the gas velocity.
A plot of the above relations between the two quantities for both subsonic and supersonic motions is shown in Fig.~\ref{fig:figure09}. There, the velocity is scaled by the sound speed $c_{\mathrm{s},2}$, based on $T_{\mathrm{sh}}$ (red curve) or $T_{\mathrm{st}}$ (black). In these units, the maximal values of $\varv/c_{\mathrm{s},2}$ for $\gamma_{\textsc{p}}=5/3$ are $\sqrt{16/5}\approx 1.78$ based on the Rankine--Hugoniot condition and $\sqrt{3}\approx 1.73$ for the stagnation point (dashed horizontal lines in  Fig.~\ref{fig:figure09}).

We first address the highest temperature gas, which should be associated with the compressed or shock-heated regions. Our temperature map in Fig.~\ref{fig:figure08} suggests that $T\gtrsim 20~\mathrm{keV}$ in some regions (for various independent analyses of the X-ray data, see \citealt{Gitti2004}, \citealt{Ota2008}, \citealt{Kreisch2016}), which immediately implies that uncertainties on the temperature based on {\it Chandra} or {\it XMM-Newton} data are very large, especially on the upper side of the confidence interval. Better constraints are provided by the {\it Suzaku} satellite, which uses a combination of X-ray CCDs with an additional HXD instrument, sensitive to temperatures above 10~keV. Using {\it Suzaku}, \cite{Ota2008} found a hot component in the SE region with $T\sim 25~\mathrm{keV}$. We note here that \cite{Ueda2018} found a higher temperature, $T\sim 29~\mathrm{keV}$, for the hotter component in the SE region, although their procedure of fixing the temperature and the contribution of the ``ambient'' component, while scaling the model normalization only by the area of the region, to derive the ``excess emission'' may bias the temperature high, since some of the volume along the line of sight is occupied by the hotter component. We have done several experiments by letting the normalization of the ambient component be free, which yields $T\sim 23~\mathrm{keV}$. We emphasize again that the uncertainties of measuring the temperature of $>20~\mathrm{keV}$ plasma with {\it Chandra} or {\it XMM-Newton} are very large. We, therefore, use only the results of {\it Suzaku}, $T=25.3\substack{+6.1\\-4.5}~\mathrm{keV}$ \citep[for systematic uncertainties, see][]{Ota2008}.

Considerable uncertainty is associated with the choice of $T_1$, even though it is easier to measure lower temperatures with {\it Chandra}. Indeed, we have several options to choose from. For instance, we can simply use the radial temperature profile shown in Fig.~\ref{fig:figure06}. At the distance of the south-eastern excess from the X-ray peak, the corresponding temperature is $\sim 17~\mathrm{keV}$. \cite{Ueda2018} found a similar temperature of $\sim 17.8~\mathrm{keV}$ for the gas north-east of the excess region (ahead in their merger scenario; see region 2a in their Table 1). These values could be affected by the complicated temperature structure of the cluster. Alternatively, one can use a mass-temperature relation based on the lensing measurements to estimate $T_1$. To this end, we use the scaling relation from \cite{Vikhlinin2009}, assuming that $M_{500}\approx 0.63 M_{200}$. For the mass of RX~J1347.5--1145 $M_{200}\sim 1.5\times10^{15}\,h^{-1}\;M_\odot$ \citep{Lu2010,Verdugo2012}, the corresponding temperature is $\sim 12~\mathrm{keV}$. However, if we use only the mass of the main subcluster $\sim 0.72 \times 10^{15}~h^{-1}~M_{\odot}$ \citep{Ueda2018}, we get $\sim7.6~\mathrm{keV}$. The latter value appears low, as the entire system is found to be permeated by higher gas temperatures, while the $12~\mathrm{keV}$ gas can be observed in several places across the cluster. Therefore, we assume rather arbitrarily that $T_1$ is somewhere between $12$ and $17~\mathrm{keV}$ and ignore further uncertainties associated with it. Corresponding temperature ratios are shown in Fig.~\ref{fig:figure09} with the vertical lines. For $T_1\sim 17~\mathrm{keV}$, there are solutions that do not involve supersonic motions; the lowest velocity that can lead to $T_{\mathrm{sh}}$ at the lowest end allowed by {\it Suzaku} is $\sim 2000~\mathrm{km}~\mathrm{s}^{-1}$. On the other hand, for $T_1\sim 12~\mathrm{keV}$ the temperature ratio is larger than $\sim$2.6, implying that the gas velocities exceed $\sim 4200~\mathrm{km}~\mathrm{s}^{-1}$.

Given that there are additional uncertainties associated with the temperature measurements \citep[see e.g.][]{Ota2008}, it is clear that neither subsonic, nor supersonic solutions, can be excluded by the X-ray temperature information alone.

\section{Conclusions}\label{sec:conclusions}
ALMA now provides an unprecedented high angular resolution view of the thermal SZ effect from galaxy clusters, and will play an invaluable role in forthcoming studies of the physics and thermodynamics of the ICM. However, as shown in Figure~\ref{fig:figure03}, the largest angular scale sampled by ALMA/ACA 90~GHz observations is limited to $\lesssim70~\mathrm{arcsec}$, and single-dish SZ measurements are crucial to complement the data and improve the spatial dynamic range to more fully sample the extent of the SZ signal.

In this work, we present a combined approach for the analysis of heterogeneous measurements through simultaneous modelling of single-dish and interferometric observations of the SZ effect. The applicability of the joint image-visibility technique is demonstrated by modelling a mixture of single-dish and interferometric observations of the well-known galaxy cluster RX~J1347.5--1145. We here briefly summarize the central results of our work.

\begin{enumerate}
    \item The combined analysis of ALMA, ACA, Bolocam, and \textit{Planck} data has been crucial for probing the pressure profile of RX~J1347.5--1145 over a wide range of spatial scales and, therefore, for deriving a comprehensive reconstruction of its thermodynamic properties. Simultaneously, it has allowed us to fully exploit the resolution and compact source sensitivity of an interferometer for modelling and removing astronomical source contamination.\\
    \item The global pressure distributions inferred from the X-ray analysis and from the joint SZ modelling, when constrained to be centred about the X-ray centroid, are in good agreement out to $\sim$1 Mpc (see Fig.~\ref{fig:figure06}).
    Along with providing validation for our modelling technique, this confirms the cool-core nature of the central region of this cluster as derived in previous independent analyses. Further, consistent with previous works, the imaging of the model-subtracted visibilities shows the presence, south-east of the X-ray peak, of a region overpressurized with respect to the cluster cool-core model. This has been generally identified as shock-heated gas.\\
    \item On the other hand, the reconstruction of the global pressure profile using no prior information on the geometry of the cluster has shown that a smooth, ellipsoidal pressure model, with centroid falling between the two BCGs, is able to describe the observed SZ signal. In this case, there is no strong evidence of shock-induced perturbations in the pressure distribution. This suggests the pressure distribution may be less disturbed than previously inferred from either the sole X-ray analysis or our X-ray-motivated SZ model. However, while no significant residual is apparent after subtraction of the best-fitting model, it is impossible to entirely rule out the presence of a shock discontinuity in the thermal pressure.\\
    \item By investigating the thermodynamic properties of RX~J1347.5--1145, we find that the south-eastern substructure seen in the X-ray image is predominantly due to isobaric rather than adiabatic perturbations. Presumably, these perturbations are related to gas stripped away from the infalling subcluster during its passage through the cluster ICM. As no strong perturbations in the pressure distribution should be expected, this is consistent with the lack of significant residuals in the SZ map after subtracting the best-fitting ellipsoidal model. Further, this alleviates the need for highly supersonic velocities required to explain the south-eastern excess as entirely due to shock-induced gas compression. However, the analysis of the gas temperature distribution inferred from X-ray data cannot unambiguously differentiate between the possible subsonic or supersonic nature of the infall of the subcluster. Further, adiabatically compressed gas is still observed ahead of the southern isobaric region.
\end{enumerate}

Future, more sensitive SZ data spanning a broader range of spatial scales will be required to conclusively measure or constrain any merger-induced pressure discontinuities, while deeper multiband kinematic SZ and X-ray micro-calorimetric data could test the assumption that the gas motion is predominantly in the plane of the sky. Deeper multiband SZ observations could also constrain the hottest ICM temperatures, which are out of reach for current X-ray instruments, through measurements of the distortion in the thermal SZ due to relativistic corrections. Given ALMA's limited ability to probe scales larger than an arcminute at frequencies $\gtrsim 100$~GHz, kinematic and relativistic SZ constraints will require improved single-dish photometric SZ imaging. The eventual extension of the modelling method to the combined reconstruction of both the SZ signal and X-ray emission will further improve the modelling of the thermodynamics of galaxy clusters, as well as provide insights into the internal structure of the ICM, for example, gas clumpiness, line-of-sight extent, and turbulence. Furthermore, this will allow for the proper treatment of the relativistic corrections to the SZ effect.

In the near future, the possibility of deeper thermal SZ observations with ALMA Band 1 (35-51~GHz) and Band 2 (67-90~GHz; see e.g. \citealt{Fuller2016, Mroczkowski_alma_memo_605}) and new thermal and kinematic SZ imaging possibilities with bolometric/photometric arrays such as TolTEC on the 50m Large Millimeter Telescope \citep{Bryan2018}, MUSTANG-2 on the 100m Green Bank Telescope \citep{Dicker2014}, and NIKA2 on the IRAM 30-meter telescope \citep{Adam2018a} will deliver such data through targeted cluster observations. However, as discussed in \cite{Mroczkowski2019}, these are often limited to $4-6~\mathrm{arcmin}$ scales, while upgrades of ALMA and the ACA to include Band 1 will only improve its largest angular scale by a factor of $2.25\times$ (to $\sim2.6~\mathrm{arcmin}$). In the longer term, a much larger field of view ($\gtrsim 1 \rm degree$), high spatial and spectral resolution submillimetre facility such as the 50-meter Atacama Large Aperture Submillimeter Telescope\footnote{AtLAST; \url{http://atlast-telescope.org/}} (see e.g. \citealt{Bertoldi2018}, \citealt{Mroczkowski2019wp}) or the Large Submillimeter Telescope \citep{Kawabe2016} would provide imaging of not only clusters, but also much larger areas of the sky, down to the sensitivities required to measure systems in the group and galaxy mass regime as well as the surrounding intergalactic filamentary structure.

\section*{Acknowledgements}
We thank the anonymous referees for providing valuable comments and suggestions, which led to the improvements in this manuscript, particularly regarding the Bayesian methods employed and the robustness of the astrophysical interpretation. The authors would also like to express special thanks to Rashid Sunyaev, Eiichiro Komatsu, Tetsu Kitayama, and Fabrizia Guglielmetti for many useful and fruitful discussions, and to thank Adi Zitrin for sharing the lensing $\kappa$ map. EC acknowledges partial support from the Russian Science Foundation grant 19-12-00369.

This paper makes use of the following ALMA data: ADS/JAO.ALMA\#2013.1.00246.S. ALMA is a partnership of ESO (representing its member states), NSF (USA) and NINS (Japan), together with NRC (Canada), MOST and ASIAA (Taiwan), and KASI (Republic of Korea), in cooperation with the Republic of Chile. The Joint ALMA Observatory is operated by ESO, AUI/NRAO and NAOJ.
This research has made use of Bolocam and Planck data hosted on the NASA/ IPAC Infrared Science Archive, which is operated by the Jet Propulsion Laboratory, California Institute of Technology, under contract with the National Aeronautics and Space Administration.
The scientific results reported in this article are based in part on data obtained from the Chandra Data Archive.
This research has made use of software provided by the Chandra X-ray Center (CXC) in the application packages \textsc{ciao}, \textsc{ChIPS}, and \textsc{Sherpa}.

\bibliographystyle{mnras}
\bibliography{rxj1347}

\appendix
\section{Joint image-visibility analysis}\label{app:analysis}
In this appendix, we present the salient features of the joint image-visibility analysis employed in this work. As already stated in the paper, the flexibility of the modelling technique makes it extensible to SZ data from other instruments, as well as to non-SZ interferometric, bolometric, photometric, and spectroscopic observations from the radio to the mm/submm regime.

\subsection{Model description}\label{sapp:model}
Suppose we obtain a data set $\bm{d}$, which provides a measure of the true sky/astronomical signal $\bm{s}$. In the case of a real experiment, this consists only of a filtered view of the real sky, due to the instrumental response and any pre-processing step applied to the data. Moreover, any measurement is inevitably contaminated by experimental noise. Assuming this to be characterized only by an additive component $\bm{n}$, the data set $\bm{d}$ can be written as
\begin{equation}
  \bm{d} = \mathbfss{T}\,\bm{s}+\bm{n}\,.
  \label{eq:data1}
\end{equation}
where the transfer operator $\mathbfss{T}$ is introduced to account for any instrument-specific filtering effects.

The description of the sky signal is of course independent of the specific observations employed in the modelling process, and any instrument-specific effect enters the overall model $\bm{m}$ only through the transfer operator $\mathbfss{T}$. We now focus on describing the signal $\bm{s}$, considered as the surface brightness distribution in a collection of directions on the sky, and introduce a model for the main astrophysical components thar are dominant when observing galaxy clusters at millimetre wavelengths.

\subsubsection{Thermal Sunyaev--Zeldovich effect}\label{ssapp:tsz}
The thermal SZ effect is the result of the inverse Compton scattering of CMB photons off the electrons in thermal motion within the hot ICM \citep[see e.g.][]{Birkinshaw1999,Carlstrom2002,Mroczkowski2019}. The variation imprinted in the CMB surface brightness in the direction of a galaxy cluster is
\begin{equation}
    \delta i_{\mathrm{t\textsc{sz}}}(\nu) = i_{\mathrm{\textsc{cmb}}} \, g_{\mathrm{t\textsc{sz}}}(\nu) \, [1+\delta_{\mathrm{t\textsc{sz}}}(\nu,T_{\mathrm{e}})] \, \vary \,,
    \label{eq:tsz1}
\end{equation}
where $i_{\textsc{cmb}}=2(k_{\mathrm{\textsc{b}}}T_{\mathrm{\textsc{cmb}}})^3/(hc)^2$, $g_{\mathrm{t\textsc{sz}}}(\nu)$ is the frequency dependence of the thermal SZ signal with amplitude $\vary$ and corrected for relativistic effects through the term $\delta_{\mathrm{e}}(\nu,T_{\mathrm{e}})$ (see below). Here, $T_{\mathrm{\textsc{cmb}}}$, $k_{\mathrm{\textsc{b}}}$, $h$, and $c$ are the average CMB temperature, the Boltzmann and Planck constants, and the speed of light, respectively.

The frequency scaling of the thermal SZ effect is derived by \citet{zeldovich1969} from the Kompaneets equation \citep{Kompaneets1957} by considering the interaction of a blackbody radiation field with a population of non-relativistic thermalized electrons,
\begin{equation}
    g_{\mathrm{t\textsc{sz}}}(\nu) = \frac{x^4\mathrm{e}^x}{(\mathrm{e}^x-1)^2}\left(x\frac{\mathrm{e}^x+1}{\mathrm{e}^x-1}-4\right) \,.
    \label{eq:tsz2}
\end{equation}
where we defined the dimensionless frequency $x=h\nu/k_{\mathrm{\textsc{b}}}T_{\mathrm{\textsc{cmb}}}$. Corrections to the thermal SZ spectrum in order to account for the presence of mildly relativistic electrons at temperature $T_{\mathrm{e}}$ can be introduced in terms of the function $\delta_{\mathrm{t\textsc{sz}}}(\nu,T_{\mathrm{e}})$, equal to the high-order solution of the Kompaneets equation \citep{Challinor1998,Itoh1998,Sazonov1998}. For a summary of the relevant relativistic correction terms, see \citet{Mroczkowski2019} and references therein.

The thermal Compton parameter $\vary$ measures the magnitude of the CMB spectral distortion due to the SZ effect, and is proportional to the line-of-sight integral of the thermal electron pressure $P_{\mathrm{e}}$ from the observer to the last scattering surface,
\begin{equation}
    \vary = \frac{\sigma_{\mathrm{\textsc{t}}}}{m_{\mathrm{e}}c^2}\int P_{\mathrm{e}}\mathrm{d}l \,.
    \label{eq:tsz3}
\end{equation}
Here, $\mathrm{d}l$ is the coordinate in the direction of the line of sight, while the constants $\sigma_{\mathrm{\textsc{t}}}$ and $m_{\mathrm{e}}$ denote the Thomson cross-section and the electron rest mass, respectively.

Departures from the Maxwellian velocity distribution of electrons, that is, due to bulk motion or non-thermal components, can induce additional spectral distortions of the CMB. However, these are generally subdominant to the thermal SZ effect \citep[e.g.][]{Mroczkowski2019}. Therefore, we decided to model only the thermal SZ signal. We do, however, consider their possible contribution when discussing the results of the fitting.

\subsubsection{Unresolved sources}\label{ssapp:ps}
Strong contamination may arise in the measured SZ effect due to the presence of point-like sources in the direction of the observed galaxy cluster. These may be due to radio emission from the cluster member galaxies, as well as foreground or background compact objects, which are bright at millimetre and submillimetre wavelengths. On the other hand, the presence of an underlying SZ component may bias the estimate of the source flux and spectral properties, thus affecting its removal from the analysed data.

When modelling an unresolved source, the general approach consists in exploiting the scale separation between the extended SZ effect component and the contaminating source, which is supposed to dominate the overall signal at the smallest scales. This is particularly suitable when analysing interferometric data, for which it is possible to take advantage of the natural spatial scale filtering for discriminating between the extended and point-like components. However, the choice of the scale range over which the fitting should be performed is fairly arbitrary. In order to avoid such freedom and, thus, possible misinterpretations of the results of the fit, any observed point-like objects are more consistently modelled jointly with the extended SZ signal.

Being point-like on the scales probed by the observations, an unresolved source can be described as a Dirac delta function at a position on the sky $\bm{x}_{\mathrm{\textsc{ps}}}=(x_{\mathrm{\textsc{ps}}},y_{\mathrm{\textsc{ps}}})$ with flux density $i_{\mathrm{\textsc{ps}i}}$ at a given reference frequency $\nu_{\mathrm{\textsc{ps}}}$,
\begin{equation}
  i_{\mathrm{\textsc{ps}}}(\bm{x},\nu) = \delta_{\textsc{d}}(\bm{x}-\bm{x}_{\mathrm{\textsc{ps}}}) \, i_{\mathrm{\textsc{ps}i}} \, g_{\mathrm{\textsc{ps}}}(\nu) \,.
  \label{eq:point1}
\end{equation}
where $g_{\mathrm{\textsc{ps}}}(\nu)$ represents the frequency spectrum of our point-source model. In our study, we adopt a simple power-law dependence with spectral index $\alpha_{\mathrm{\textsc{ps}}}$, yielding
\begin{equation}
    g_{\mathrm{\textsc{ps}}}(\nu) = (\nu/\nu_{\mathrm{\textsc{ps}}})^{\alpha_{\mathrm{\textsc{ps}}}}.
    \label{eq:point2}
\end{equation}

\subsubsection{Other components}\label{ssapp:other}
The above procedure only makes sense for those objects characterized by a flux density above the threshold set by the instrument sensitivity (or, potentially, by the confusion limit of the observations if not enough constraints are available to model it). On the other hand, faint, undetected radio sources identified using ancillary radio catalogues may be subtracted a priori from the input data sets if their fluxes can be accurately estimated. However, uncertainties in the spectral model may result in the wrong extrapolation of the source fluxes at the required frequencies. Moreover, residual unresolved sources near or below the confusion limit may still contribute to the total measured signal and affect the analysis of the SZ effect by introducing significant contamination. It is then possible to account for the resulting bias in terms of additional uncertainties in the measured flux density.

Similarly, the amplitude of the primary CMB anisotropies at scales comparable to the angular size of a given galaxy cluster may be non-negligible with respect to the thermal SZ flux density. A statistical description of the overall impact of the CMB on the measured SZ signal can be derived from the power spectrum of CMB fluctuations.

We include the overall effect of both point-like sources and CMB contamination in our noise model. In practice, the covariance matrix employed in the computation of the likelihood function of our model can be generalized to
\begin{equation}
    \mathbfss{C} = \mathbfss{C}_{\textsc{n}}+\mathbfss{C}_{\mathrm{\textsc{ps}}}+\mathbfss{C}_{\mathrm{\textsc{cmb}}}
    \label{eq:gencov}
\end{equation}
Here, $\mathbfss{C}_{\textsc{n}}$, $\mathbfss{C}_{\mathrm{\textsc{ps}}}$, and $\mathbfss{C}_{\mathrm{\textsc{cmb}}}$ define the instrumental, unresolved-source, and CMB covariance matrices, respectively. For simplicity, we assumed that neither the compact sources nor CMB are spatially correlated with the SZ signal.

\subsection{Data likelihood}\label{sapp:datalike}
\begin{figure*}
  \begin{center}
  \includegraphics[width=\textwidth]{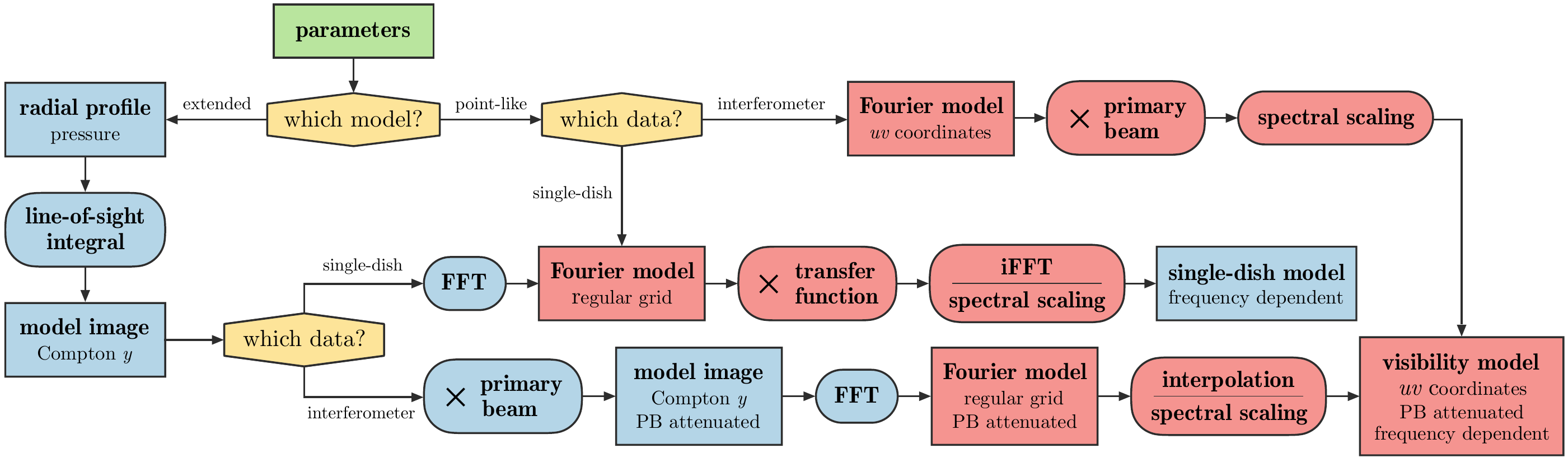}
  \end{center}
  \caption{Schematic flow diagram of our algorithm for modelling single-dish and interferometric data. The solid rectangles represent the data products employed in our computations, while the specific operations are indicated in rounded rectangles. We highlight in blue and in red all the quantities and functions defined respectively in image- and Fourier-space.}
  \label{fig:figure02}
\end{figure*}

As shown in Eq.~\eqref{eq:data1}, the proper comparison of a parametric representation of an astrophysical signal to any input measurement relies on the characterization of the effects of the transfer function on the observed sky. Here, we detail how they can be taken into account differently in the case of single-dish and interferometric observations. A schematic representation of the derivation of the data models is presented in Figure~\ref{fig:figure02}.
For the sake of readability, we omit hereafter the subscript denoting the specific subset of data and the respective model unless strictly required. All the quantities discussed here can be generalized to each of the individual observations.

\subsubsection{Image-space observations}\label{sssec:imspace}
For a set of image-domain data, the effect of the transfer operator in Eq.~\eqref{eq:data1} can be summarized in terms of a combination of the smoothing due to the finite angular resolution of the telescope, and some large-scale filtering as a consequence of the scanning and map-making strategies. It acts on the true signal as a convolution kernel, making it computationally more convenient to be treated in Fourier space. By applying the convolution theorem, the map of a generic model component can be written as
\begin{equation}
    \bm{m}_{\mathrm{img}}(\bm{x},\nu) = [\mathbfss{T}_{\mathrm{img}}*\bm{s}] (\bm{x},\nu) = \int [\tilde{\mathbfss{T}}_{\mathrm{img}}\,\tilde{\bm{s}}](\bm{u},\nu) \, \mathrm{e}^{-2\pi i \bm{u}\cdot\bm{x}} \mathrm{d}\bm{u} \,,
    \label{eq:data2}
\end{equation}
where we denote with a tilde the Fourier transform of a given function. As before, $\bm{x}$ represents the vectorial direction on the sky of the observed signal, while $\bm{u}$ describes the Fourier-space coordinates. 

The unfiltered model for the extended thermal SZ signal over a set of directions $\bm{x}$ is computed as
\begin{equation}
  \bm{s}_{\mathrm{t\textsc{sz}}}(\bm{x},\omega) = i_{\textsc{cmb}} \, \bar{g}_{\mathrm{t\textsc{sz}}}(\omega,T_{\mathrm{e}}) \, \vary(\bm{x}) \,,
  \label{eq:model1}
\end{equation}
where we introduced the bandpass function $\omega = \omega(\nu)$. Indeed, a real instrument observes the sky over a given range of frequencies with non-uniform spectral response, due to a combination of instrumental effects and atmospheric transmission. The frequency dependence of the measured SZ effect is then expressed by the bandpass-averaged spectral function
\begin{equation}
  \bar{g}_{\mathrm{t\textsc{sz}}}(\omega,T_{\mathrm{e}}) = \frac{\int \omega(\nu') \, g_{\mathrm{t\textsc{sz}}}(\nu')[1+\delta_{\mathrm{t\textsc{sz}}}(\nu',T_{\mathrm{e}})] \, \mathrm{d}\nu'}{\int \omega(\nu') \, \mathrm{d}\nu'}  \,,
  \label{eq:freq1}
\end{equation}

In the specific case of a point-like source, we can employ the sifting property of the Dirac delta function to immediately define its model as being equal to the convolution kernel $\mathbfss{T}_{\mathrm{img}}$ centred at the position of the source and scaled by the source amplitude at the frequency $\nu$ of the given observation. From Eq.~\eqref{eq:point1}, considering bandpass-averaged quantities,
\begin{equation}
    [\mathbfss{T}_{\mathrm{img}}\ast\bm{s}_{\mathrm{\textsc{ps}}}](\bm{x},\omega) = \bar{g}_{\mathrm{\textsc{ps}}}(\omega) \, \bm{s}_{\mathrm{\textsc{ps}}}(\bm{x}_{\mathrm{\textsc{ps}}}) \, \mathbfss{T}_{\mathrm{img}}(\bm{x}-\bm{x}_{\mathrm{\textsc{ps}}},\omega) \,.
    \label{eq:point3}
\end{equation}
where $\bar{g}_{\mathrm{\textsc{ps}}}$ is defined analogously to Eq.~\eqref{eq:freq1} for the case of a power-law spectral model. By using the Fourier shift theorem, the transfer term on the right-hand side of the above equation can then be computed as
\begin{equation}
    \mathbfss{T}_{\mathrm{img}}(\bm{x}-\bm{x}_{\mathrm{\textsc{ps}}},\omega) = \int \left[\tilde{\mathbfss{T}}_{\mathrm{img}}(\bm{u},\omega) \, \mathrm{e}^{2\pi i \bm{u}\cdot\bm{x_{\mathrm{\textsc{ps}}}}}\right] \mathrm{e}^{-2\pi i \bm{u}\cdot\bm{x}} \mathrm{d}\bm{u} \,.
    \label{eq:point4}
\end{equation}

The total model map $\bm{m}_{\mathrm{img}}$ employed for evaluating the image likelihood function can then be easily obtained by summing over all the arbitrary number of components of our parametric model.

The generalized covariance matrix required for computing the data likelihood may be strongly non-diagonal in the case of image-space data (for a discussion, we refer to e.g. \citealt{Condon1974}, and \citealt{Knox2004}). Even omitting the effects of gravitational lensing and of the spatial clustering of the sources, the component of the covariance matrix associated with the confusion noise shows off-diagonal terms due to the spatial correlation induced by the finite resolution. On the other hand, from the properties of the spherical harmonics, the elements of the CMB covariance matrix associated with any two pixels at a given angular separation $\theta_{\mathrm{ij}}$ can be written as
\begin{equation}
    [\mathbfss{C}_{{\textsc{cmb}}}]_{\mathrm{ij}} = \sum_{\ell}\frac{2l+1}{4\pi} \, C_{\ell} \, P_{\ell}(\cos{\theta_{\mathrm{ij}}}) \,,
    \label{eq:cov1}
\end{equation}
where $C_{\ell}$ is the value at the multipole $l$ of the primary CMB power spectrum and $P_l$ the Legendre polynomial of order $\ell$.

\subsubsection{Interferometric data}\label{ssapp:uvspace}
The observation of a given signal in direction $\bm{x}$ on the sky performed by a single baseline of a radio interferometer at a frequency $\nu$ can be modelled as \citep{Thompson1986}
\begin{align}
    \bm{V}(\bm{u},w,\nu) = &\int \frac{1}{\sqrt{1-|\bm{x}-\bm{x}_0|^2}}\, \mathbfss{A}(\bm{x},\nu) \, \bm{s}(\bm{x},\nu) \notag\\
                           &\times \mathrm{e}^{-2\pi i \big[\bm{u}\cdot(\bm{x}-\bm{x}_0)+w\big(\sqrt{1-|\bm{x}-\bm{x}_0|^2}-1\big)\big]} \, \mathrm{d}\bm{x} \,,
\end{align}
where $\bm{V}(\bm{u},w,\nu)$ is commonly referenced as visibility function. Here, $\bm{u}$ is the two-dimensional Fourier space coordinate vector --- position in the so-called $uv$-plane --- which represents the spatial wavelengths corresponding to the projection of the baselines on the plane of the sky, while $w$ is the baseline vector component parallel to the line of sight and measured in the direction of the phase reference position $\bm{x}_0$. The function $\mathbfss{A}(\bm{x},\nu)$ describes the attenuation of the sky map in a given direction $\bm{x}$ induced by the non-uniform beam response pattern of an antenna, and is a function of frequency. In general, the primary beam function vanishes rapidly with the distance from the phase centre, so that the resulting single-pointing field of view is limited mainly to the region where $|\bm{x}-\bm{x}_0|\ll1$. This further implies that it is possible to approximate the observed surface brightness as being distributed on the plane tangent to the sky at the position $\bm{x}_0$. In such case, $w=0$ and the position-dependent denominator can be approximated by unity. Therefore, we can simplify the visibility function as follows:
\begin{equation}
    \bm{V}(\bm{u},\nu) \approx \bm{V}(\bm{u},0,\nu) = \int \mathbfss{A}(\bm{x},\nu) \, \bm{s}(\bm{x},\nu) \, \mathrm{e}^{-2\pi i \bm{u}\cdot{x}}\mathrm{d}\bm{x} \,.
\end{equation}
This means that, in the limit of such flat-sky approximation, an interferometric measurement can be directly related to the Fourier transform of the sky surface brightness attenuated by the antenna response pattern. We note that, while the primary beam sets the field of view of an interferometer, and must be accounted for to correct the amplitude of a source within this field, the case differs for single-dish data, where the primary beam of the instrument defines its resolution element. Therefore, for calibrated single-dish observations, no correction is required other than that for the transfer function, which includes the low-pass-filtering effect of the beam itself (see Sec.~\ref{sssec:imspace}).

Finally, we can account for the overall, but still incomplete sampling of the visibility space performed by the whole ensemble of baselines comprising a radio interferometer by introducing an operator $\tilde{\mathbfss{T}}_{\mathrm{vis}}(\bm{u},\nu)$. Using the same notation as the previous sections, the model for the interferometric data can then be defined as
\begin{equation}
    \tilde{\bm{m}}_{\mathrm{vis}}(\bm{u},\nu) = \tilde{\mathbfss{T}}_{\mathrm{vis}}(\bm{u},\nu) \int  \mathbfss{A}(\bm{x},\nu) \, \bm{s}(\bm{x},\nu) \, \mathrm{e}^{2\pi i \bm{u}\cdot\bm{x}} \mathrm{d}\bm{x} \,.
    \label{eq:data3}
\end{equation}

The visibilities for extended model components are computed by resampling the regularly gridded Fourier transform of the corresponding model map onto the coordinates of the sparse interferometric data.
The input, unsmoothed model image is produced as for the case of the image-domain fitting described earlier. However, instead of applying any filtering, it is only attenuated by the input primary beam pattern before being Fourier transformed.

Unlike for extended sources, the modelling of unresolved sources is straightforward, since, as mentioned in the previous section, they can immediately be defined in Fourier space. Substituting the signal model in Eq.~\eqref{eq:data3} with the expression for a point-like component of Eq.~\eqref{eq:point1}, we obtain
\begin{equation}
  [\tilde{\bm{m}}_{\mathrm{vis}}(\bm{u},\omega)]_{\mathrm{\textsc{ps}}} = \tilde{\mathbfss{T}}_{\mathrm{vis}}(\bm{u},\omega) \mathbfss{A}(\bm{x}_{\mathrm{\textsc{ps}}},\omega) \, i_{\mathrm{\textsc{ps}i}} \, \bar{g}_{\mathrm{\textsc{ps}}}(\omega) \, \mathrm{e}^{2\pi i \bm{u}\cdot\bm{x}_{\mathrm{\textsc{ps}}}} \,.
  \label{eq:point5}
\end{equation}
It follows that any residual unresolved sources result in an offset term in the complex amplitude of the interferometric data. Therefore, we can avoid including the confusion term inside the generalized noise covariance matrix and treat it as a constant level in the Fourier-space model to be subtracted during the parametric reconstruction.

On the other hand, CMB power declines significantly when considering ever smaller scales as a consequence of Silk diffusion damping \citep{Silk1968}, implying that CMB contamination affects mainly the visibilities with the shortest spacings. Moreover, since it represents the effect of a spatially defined astrophysical signal, it generates correlated visibilities and, then, a non-diagonal covariance matrix. The computation of the CMB covariance for interferometric observations is presented in \citet{Hobson2002b}. However, the filtering effect resulting from the missing short spacings can make the contribution to noise from CMB fluctuations generally negligible even at large scales when compared to the instrumental component. This is assumed to be uncorrelated, leading to a diagonal noise covariance matrix. It follows that, in such a case, the exponent of the interferometric likelihood reduces to
\begin{equation}
    \big[(\bm{d} - \bm{m})^{\dagger} \mathbfss{C}^{-1} (\bm{d} - \bm{m})\big]_{\mathrm{vis}} \approx \tilde{\bm{w}}_{\mathrm{vis}}\cdot|\tilde{\bm{d}}_{\mathrm{vis}}-\tilde{\bm{m}}_{\mathrm{vis}}|^2 \,,
    \label{eq:chisq2}
\end{equation}
where the factor $\tilde{\bm{w}}_{\mathrm{vis}}$ represents the set of theoretical post-calibration visibility weights, equal to the inverse of the noise variance of the corresponding interferometric measurement \citep{Wrobel1999}. Here, the dagger symbol is used to denote the conjugate transpose.

\subsubsection{Integrated Compton parameter}\label{ssapp:integraly}
Additional constraints on the pressure profile may be introduced by including information about the integrated thermal SZ flux obtained from the aperture photometry of the cluster Compton $\vary$ map. This is proportional to the volume integral of the ICM pressure distribution and, hence, it is a fundamental proxy of the total thermal energy and mass of a galaxy cluster \citep[see e.g.][]{Mroczkowski2019}.

We can define the cylindrically integrated Compton parameter $Y_{\mathrm{cyl}}$ as the surface integral over a given solid angle $\Omega_{\mathrm{max}}$ of the cluster Compton parameter $\vary$ distribution,
\begin{equation}
    Y_{\mathrm{cyl}} = \int_{\Omega_{\mathrm{max}}} \!\! \vary \,\mathrm{d}\Omega \,.
\end{equation}
In our analysis, instead of characterising possible contributions arising from considering images with given finite angular resolution and large-scale filtering properties, we compare the integrated Compton parameter $Y_{\mathrm{cyl}}$ measured from a map characterized by the transfer operator $\mathbfss{T}_{\mathrm{int}}$ with the respective value obtained from a filtered version of the model Compton $\vary$ map, $[\mathbfss{T}_{\mathrm{int}}\ast\bm{s}_{\mathrm{t\textsc{sz}}}]$. The corresponding likelihood function is then combined with the ones obtained by modelling the image and interferometric data.

\section{Triaxial ellipsoidal profile}\label{app:ellipsoid}
Any radial pressure profile can be easily extended to describe a triaxial ellipsoidal distribution. Indeed, the information about the different characteristic extents of the ellipsoid principal axes can be included in a generalized dimensionless radius $\xi$, which can then be considered instead of the ratio $(r/r_{\mathrm{s}})$ for computing the pressure profile in Eq.~\eqref{eq:pressure1}.
Since we cannot recover information about the line-of-sight geometry of the pressure distribution using SZ data only, the main assumption in our computations of the thermal SZ signal from an ellipsoidal cluster consists in considering two of the principal axes to lie on the plane of the sky and the other to be aligned with the line-of-sight direction. Furthermore, the extent of the latter is hard-coded with the assumption that it is equal to the inverse root mean square average of the plane-of-sky semi-axes. It follows
\begin{equation}
    \xi = \frac{1}{r_{\mathrm{s}}}\left( \, r^2_{\mathrm{a}} \, + \, \frac{1}{1-\varepsilon^2} \, r^2_{\mathrm{b}} \, + \, \frac{1}{2} \, \frac{2-\varepsilon^2}{1-\varepsilon^2} \, r^2_{\mathrm{c}} \, \right)^{1/2}\,,
    \label{eq:ellipsoid1}
\end{equation}
where $r_{\mathrm{a}}$, $r_{\mathrm{b}}$, and $r_{\mathrm{c}}$ are the radii measured along the three principal axes of the ellipsoid, respectively, and $\varepsilon$ is the eccentricity of the elliptical profile projected on the sky. If we denote with $\delta_0$ the position angle of the plane-of-sky major axis taken from north through east, we can write
\begin{align}
    r_{\mathrm{a}} &= (x-x_0)\,\cos{y_0}\,\cos{\delta_0}-(y-y_0)\,\sin{\delta_0}\nonumber\\
    r_{\mathrm{b}} &= (x-x_0)\,\cos{y_0}\,\sin{\delta_0}+(y-y_0)\,\cos{\delta_0}\\
    r_{\mathrm{c}} &= z-z_0\,, \nonumber
\end{align}
where $(x-x_0)$, $(y-y_0)$ and $(z-z_0)$ are differences in the right ascension, declination and line-of-sight distance of a given point with respect to the centre of the ellipsoidal profile with coordinates $(x_0,y_0,z_0)$.

\section{Data-free sampling}\label{app:datafree}

We here report the results of the data-free run of the modellings presented in Sections~\ref{sssec:sphere} and \ref{sssec:ellipse}. This corresponds to letting the sampler explore only the prior space, and can then provide fruitful insight into any effects on the parameter reconstruction introduced by the specific choice of the prior distributions. 

\begin{figure}
  \begin{center}
    \includegraphics[width=\columnwidth]{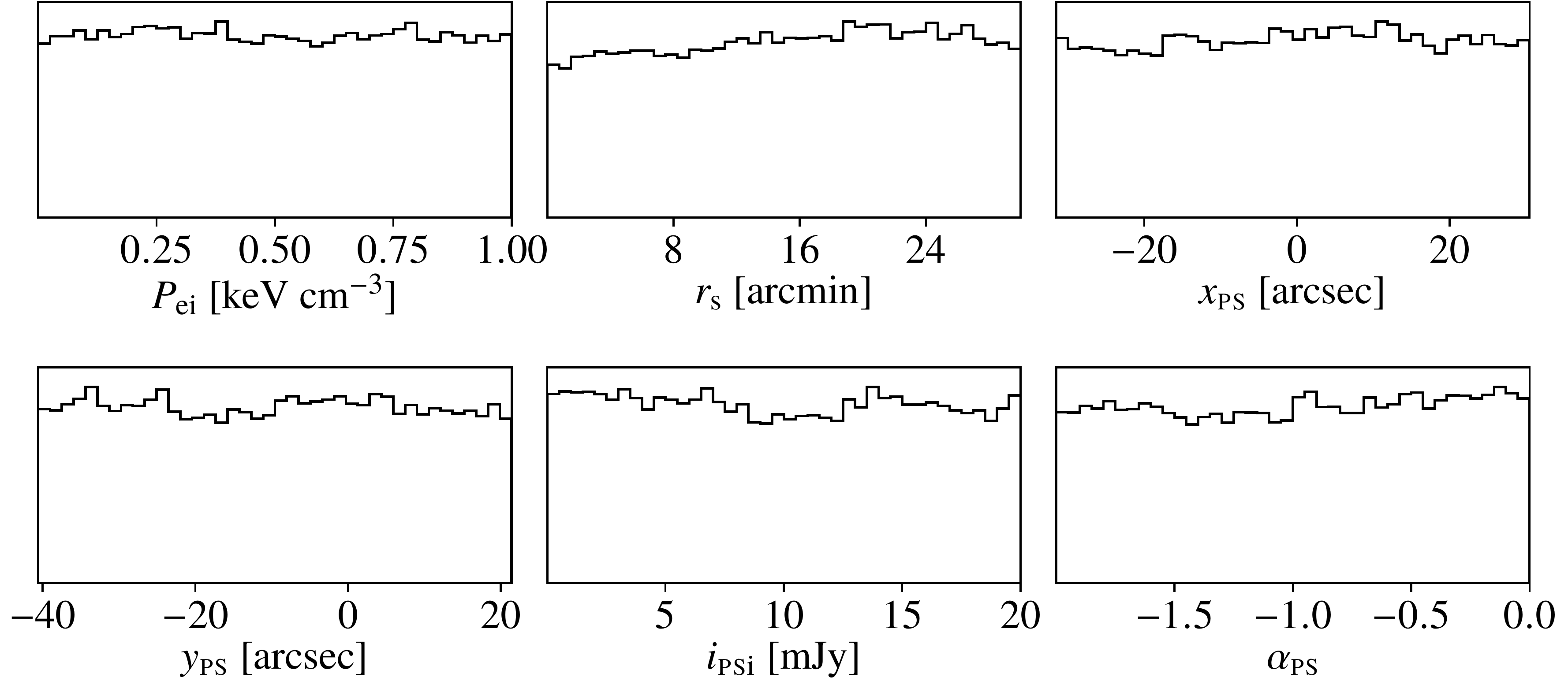}
  \end{center}
  \caption{Marginalized posterior distributions for the prior-only run for the parametric modelling of the cool-core region of RX J1347.5--1145. A description of the model parameters and the corresponding priors can be found in Section~\ref{sssec:sphere} and references therein.}
  \label{fig:figure10}
\end{figure}

\begin{figure}
  \begin{center}
    \includegraphics[width=\columnwidth]{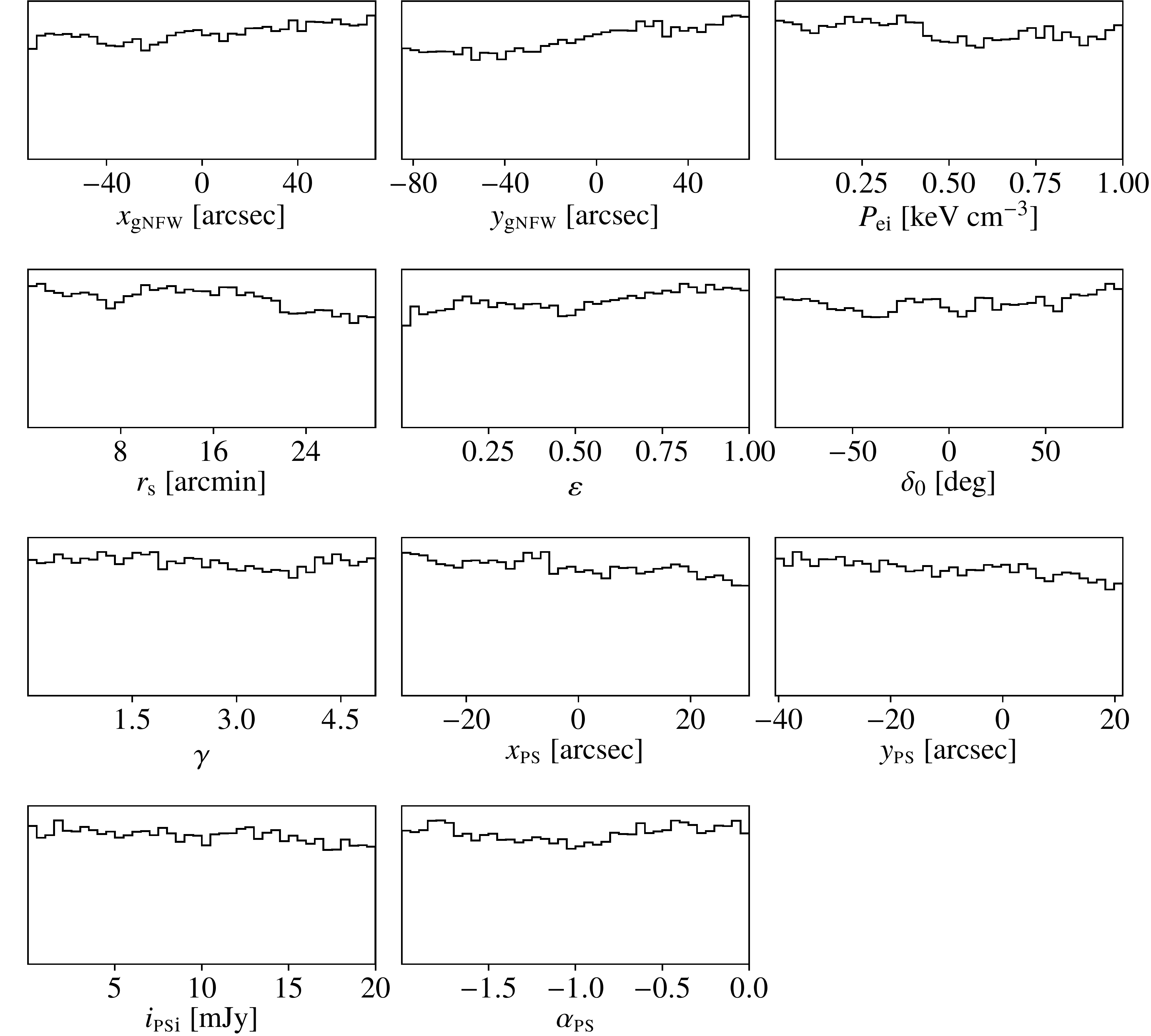}
  \end{center}
  \caption{Same of Fig.~\ref{fig:figure10} but for the ellipsoidal model of Section~\ref{sssec:ellipse}.}
  \label{fig:figure11}
\end{figure}

Both for the cool-core and global pressure profiles, all the prior distributions are recovered according to the input information.

\bsp
\label{lastpage}
\end{document}